\documentclass[a4paper,11pt]{article}

\usepackage{color}
\usepackage{ulem}
\usepackage{enumerate}
\usepackage{amsmath}
\usepackage{graphicx}
\usepackage[colorlinks=true,linkcolor=red,citecolor=blue,urlcolor=blue,bookmarks]{hyperref}
\usepackage{amsmath} 
\usepackage{amssymb}

\def \be {\begin{equation}}
\def \ee {\end{equation}}
\def \ba {\begin{array}}
\def \ea {\end{array}}
\def \bea {\begin{eqnarray}}
\def \eea {\end{eqnarray}}

\def \ble {\begin{widetext}\begin{equation}}
\def \ele {\end{equation}\end{widetext}}
\def \blea {\begin{widetext}\begin{eqnarray}}
\def \elea {\end{eqnarray}\end{widetext}}

\def \nn {\nonumber}

\def \trans {\mathcal{T}^{\psi|\phi}}
\def \transA {\mathcal{T}_A^{\psi|\phi}}

\def \blea {\begin{widetext}\begin{eqnarray}}
\def \elea {\end{eqnarray}\end{widetext}}

\def \p {\partial}

\begin{document}

\title{ Spectral Projections for Density Matrices in Quantum Field Theories}

\author{Wu-zhong Guo}


\date{}
\maketitle

\vspace{-10mm}
\begin{center}
{\it School of Physics, Huazhong University of Science and Technology,\\
Luoyu Road 1037, Wuhan, Hubei
430074, China
\vspace{1mm}
}
\vspace{10mm}
\end{center}

\begin{abstract}
In this paper, we investigate the spectral projection of density matrices in quantum field theory. With appropriate regularization, the spectral projectors of density matrices are expected to be well-defined. These projectors can be obtained using the Riesz projection formula, which allows us to compute both the density of eigenvalues and the expectation values of local operators in the projected states. We find that there are universal divergent terms in the expectation value of the stress energy tensor, where the coefficients depend universally on the density of eigenvalues and a function that describes the dependence of eigenvalues on boundary location. Using projection states, we can construct a series of new states in quantum field theories and discuss their general properties, focusing on the holographic aspects. We observe that quantum fluctuations are suppressed in the semiclassical limit. We also demonstrate that the fixed area state, previously constructed using gravitational path integrals, can be constructed by suitably superposition of appromiate amount of projection states. Additionally, we apply spectral projection to non-Hermitian operators, such as transition matrices, to obtain their eigenvalues and densities. Finally, we highlight potential applications of spectral projections, including the construction of new density and transition matrices and the understanding of superpositions of geometric states.
\end{abstract}

\maketitle

\tableofcontents
\section{Introduction}

A fundamental question in gravity research is understanding the holographic duality between a gravity system and its lower-dimensional field theory. Within the well-established AdS/CFT correspondence \cite{Maldacena:1997re,Gubser:1998bc,Witten:1998qj}, it is possible to construct exact dualities for observables. Entanglement entropy, as a fundamental probe for detecting the structure of wave functions or density matrices in the boundary field theory, is linked to the minimal surface in the gravitational side \cite{Ryu:2006bv}\cite{Hubeny:2007xt}. Motivated by this elegant formula, significant insights have been gained, providing a deeper understanding of the exact duality between the two theories \cite{VanRaamsdonk:2010pw}-\cite{Dutta:2019gen}. 

Given a quantum state with density matrix $\rho$. The reduced density matrix of a subregion $A$, denoted by $\rho_A:=tr_{\bar A}\rho$, encodes information about $A$ as well as the correlations or entanglement between different spacetime regions. Entanglement entropy (EE) $S(\rho_A):= -tr(\rho_A\log \rho_A)$ is the von Neumann entropy, which captures only partial information about $\rho_A$, The full information of an operator is encoded in its spectrum and spectral projections, which have not been extensively studied. This is partly because, in quantum field theories (QFTs), it is generally challenging to formulate spectral projections for a given operator, as it is believed that the Hilbert space is infinite-dimensional. The density matrices and spectral projections in QFTs may be far more subtle than the finite-dimensional matrices with which we are familiar \cite{Witten:2018zxz}. In QFTs, one cannot directly obtain the spectrum and spectral projections by diagonalizing $\rho_A$. A practical approach is to reconstruct the spectrum using the R\'enyi entropy, a generalization of EE. The $n$-th R\'enyi entropy is defined as $S_n(\rho_A):=\frac{\log tr\rho_A^n}{1-n}$. However, there have been studies on the spectral density of reduced density matrices in various contexts, see \cite{Li2008}-\cite{Bai:2022obp} and references therein.

In this paper, our primary aim is to formalize the spectral projection of $\rho_A$ by using Riesz projection formula. In principle the The Riesz formula is, in principle, applicable to the eigenvalues of compact operators. Without regularization, it is not expected that we can directly apply the Riesz formula to $\rho_A$. However, we anticipate that with proper regularization, including a UV cut-off, the operator $\rho_A$ will behave similarly to the finite-dimensional case. In some simple examples, we demonstrate that the eigenvalues are discrete. Using spectral projections, one can compute the density of eigenvalues and the expectation value of local operators in the projected states. We will illustrate this with examples and compare our results with previous findings obtained using different methods \cite{Calabrese2008}\cite{Guo:2020roc}.

We study the properties of the projection state in detail. We find that the expectation value of the stress energy tensor exhibits a universal divergent term near the boundary of the subsystem $A$. The coefficients of these divergent terms depend solely on the density of eigenvalues and a function that captures the eigenvalue dependence on the boundary location (see Section \ref{section_universal} for details).

Another important application of spectral projection is the construction of a new class of states in QFTs. We primarily focus on the properties of these states from the perspective of holography. We discuss the connected correlation functions in these states and find that quantum fluctuations are suppressed in the semiclassical limit. We also argue that by appropriately superposing projection states, one can construct the fixed area states, which have been previously discussed in the bulk using quantum error correction codes and the gravitational path integral \cite{Akers:2018fow}\cite{Dong:2018seb}. A key feature of fixed area states is their flat spectrum, making the R\'enyi entropy independent of the index $n$. We show that producing fixed area states involves the superposition of an appropriate number of projection states.

The spectral projection method can also be applied to non-Hermitian operators, such as the transition matrices discussed in \cite{Nakata:2020luh}. We provide examples to demonstrate how to obtain the eigenvalues and the density of eigenvalues for these non-Hermitian transition matrices—something that could not be achieved using previous methods.

The paper is organized as follows. In Section \ref{section_Riesz}, we introduce the spectral projection using the Riesz formula. We also demonstrate how to obtain the density of eigenvalues and the expectation value of local operators in projection states in Section \ref{section_density} and Section \ref{section_correlator_local}, respectively. An example in 2-dimensional CFTs is presented in Section \ref{section_example}. In Section \ref{section_universal}, we identify universal divergent terms in the expectation value of the stress-energy tensor. Section \ref{section_geometry} focuses on the holographic aspects of projection states and their superpositions. We discuss the connected correlation functions in these states in Section \ref{Fluctuations_projection}. The relationship between fixed-area states and projection states is explored in Section \ref{section_fixed}. The final section \ref{section_summary} provides a summary and discussion, highlighting potential applications of the results presented in this paper.

\section{Spectral projection of density matrix}\label{Section_basis}
In this section, we will discuss spectral projection for a given density matrix and explore the relationship between the density of eigenvalues. By using spectral projections, one can construct projection states and study their properties.

\subsection{Riesz Projection}\label{section_Riesz}
\begin{figure}
  \centering
 \includegraphics[scale=0.8]{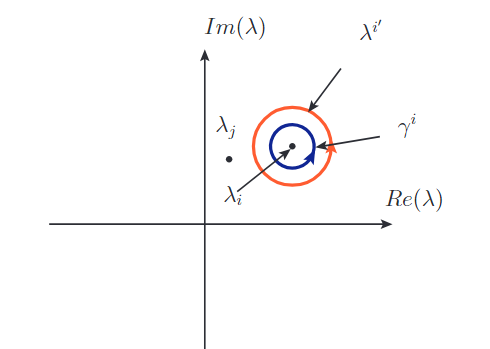}
 \caption{Contours $\gamma^i$ and $\gamma^{i'}$ surround the eigenvalue $\lambda_i$ on the complex plane of $\lambda$. The eigenvalue $\lambda_j$ represents other eigenvalues. The contours $\gamma^i$ and $\gamma^{i'}$ should only enclose $\lambda_i$.  }
\label{fig1}
\end{figure}

Let $M$ be compact operator acting on the Hilbert space $\mathcal{H}$, $\sigma(M)$ be the spectrum of $M$. The resolvent $R_M(z):=(z-M)^{-1}$ is a holomorphic operator-valued function in the set $\mathbb{C}\backslash \sigma(M)$. Assume $\{\lambda_i\}$ are the eigenvalues of $M$.  The Riesz projection $\mathbb{P}_M(\lambda)$ for the eigenvalue $\lambda_i$ can be expressed as
\bea\label{Riesz_projection}
\mathbb{P}_M(\lambda_i)=\frac{1}{2\pi i} \int_{\gamma^i}dz R_M(z),
\eea 
where $\gamma^i$ is the oriented counterclockwise contour that encloses only the point $\lambda_i$, see for example \cite{Riesz}. It can be shown $\mathbb{P}_M(\lambda_i)$ is a projection $\mathbb{P}_M(\lambda_i)^2=\mathbb{P}_M(\lambda_i)$. By definition  we have
\bea\label{projection_square}
\mathbb{P}_M(\lambda_i)^2=\frac{1}{(2\pi i)^2} \int_{\gamma^i} \int_{{\gamma^i}'}R_M(z)R_M(z')dzdz',
\eea
where $\gamma^i$ and ${\gamma^i}'$ are two contours surrounding $\lambda_i$ shown in Fig.\ref{fig1}. The resolvent $R_M(z)$ satisfies the following relation
\bea
R_M(z)R_M(z')=\frac{R_M(z')-R_M(z)}{z-z'}.
\eea
Taking the above formula to Eq.(\ref{projection_square})
\bea
&&\mathbb{P}_M(\lambda_i)^2=\frac{1}{(2\pi i)^2} \int_\gamma \int_{\gamma'}\frac{R_M(z')-R_M(z)}{z-z'} dz dz'\nn \\
&&\phantom{\mathbb{P}_M(\lambda_i)^2}=\frac{1}{(2\pi i)^2} \int_{{\gamma^i}'} dz' R_M(z') \int_{\gamma^i} dz \frac{1}{z-z'} dz \nn \\
&&\phantom{\mathbb{P}_M(\lambda_i)^2=}-\frac{1}{(2\pi i)^2} \int_{\gamma^i} dz R_M(z) \int_{{\gamma^i}'} dz' \frac{1}{z-z'} dz\nn \\
&&\phantom{\mathbb{P}_M(\lambda_i)^2}=\mathbb{P}_M(\lambda_i),
\eea
where we use $\int_{\gamma^i} dz \frac{1}{z-z'} dz=1$ and $\int_{{\gamma^i}'} dz' \frac{1}{z-z'} dz=0$. It can be further shown that 
\bea\label{projection_relation}
\mathbb{P}_M(\lambda_i)\mathbb{P}_M(\lambda_{i'})=\delta_{\lambda_i \lambda_{i'}}\mathbb{P}_M(\lambda_i).
\eea
The Riesz projection can be applied to Hermitian operators, such as a density matrix, and to non-Hermitian operators, such as a transition matrix. In the following sections, we will use examples to demonstrate the application of Riesz projections.

\subsection{Density of eigenvalues}\label{section_density}

For the density matrix $\rho$ with discrete eigenvalues $\lambda_i \in (0,\lambda_m)$ with $\lambda_m\le 1$, where $\lambda_m$ is the maximal eigenvalue.  The dimension of the degenerate subspace $\mathbb{P}_\rho(\lambda_i)$ is denoted by $\mathcal{P}(\lambda_i)$. It is obvious that $\mathcal{P}(\lambda_i):=tr\mathbb{P}(\lambda_i)$, which can be evaluated by the following formula
\bea\label{density}
\lambda_i \mathcal{P}(\lambda_i)=\frac{1}{2\pi i} \int_{\gamma^i}dz tr[\rho R_\rho(z)]=\frac{1}{2\pi i} \int_{\gamma^i}dz \sum_{n=1}^\infty \frac{tr\rho^n}{z^n}.
\eea
According to the definition of R\'enyi entropy, $tr\rho^n =e^{(1-n)S_n(\rho)}$. Thus once $S_n(\rho)$ is known, one could compute the function $\mathcal{P}(\lambda_i)$ by using formula (\ref{density}).

In previous studies the density of eigenvalues at $\lambda$ is defined as 
\bea\label{continuous}
\mathcal{P}(\lambda):=\sum_{i}\delta(\lambda_i-\lambda).
\eea
With this one could evaluate the $\mathcal{\lambda}$ by using Inverse Laplace transformation, see \cite{Calabrese2008}\cite{Guo:2020roc}. The quantity $\mathcal{P}(\lambda)$ can be evaluated by using the information from $S_n(\rho)$, which is consistent with the formula (\ref{density}).

Before discussing the properties of the projection states, let us consider the subtleties of the functions  $\mathcal{P}(\lambda_i)$ and $\mathcal{P}(\lambda)$. In the approach using Riesz projection, we assume that the spectrum of the density matrix $\rho$
is discrete. However, this assumption is subtle in QFTs. It is generally expected that the spectrum of the density matrix in QFTs should be continuous, as QFTs have an infinite Hilbert space. The density matrix and EE may not be well-defined from the perspective of algebraic QFTs \cite{Witten:2018zxz}. However, in practical computations, one can obtain a finite value of EE by introducing a UV cut-off $\epsilon$. It appears that the introduction of $\epsilon$ alters the underlying algebraic structure\footnote{Practically, one can roughly understand the regularized Hilbert space by discretizing the theory on a lattice.}. 

In the following, we will study the example of one interval in the vacuum state of 2D Conformal Field Theories (CFTs). In this example, the role of the UV cut-off $\epsilon$ is clear. As shown in \cite{Cardy:2016fqc}, in the path integral representation of states, the regularization that removes a region of thickness $\epsilon$ around the entanglement boundary can be interpreted as a projection of the Hilbert space to a smaller subspace. Through this process, one can obtain the density matrix and modular Hamiltonian by conformal mapping to a cylinder. This approach explicitly shows that the spectra of the density matrix and modular Hamiltonian are discrete. In the present paper, we will adopt this viewpoint. In the following sections, we will use examples to demonstrate that the Riesz projection can yield the correct density of eigenvalues. Moreover, the function $\mathcal{P}(\lambda_i)$ given by Eq.(\ref{density}) is same as the function defined by Eq.(\ref{continuous}) with $\lambda=\lambda_i$. In practical calculations, we encounter summations of discrete eigenvalues, such as $\sum_i f(\lambda_i)\mathcal{P}(\lambda_i)$. We also expect  that
\bea\label{continuous_integral}
\sum_i f(\lambda_i)\mathcal{P}(\lambda_i)=\int_0^{\lambda_m} d\lambda f(\lambda)\mathcal{P}(\lambda),
\eea 
in the continous limit $\epsilon\to 0$. In the following, we will use integration to evaluate the physical quantities of interest.

\subsection{Projection states and expectation value of local operators}\label{section_correlator_local}
Consider a pure state $|\psi\rangle$ in QFTs. For a  subsystem $A$ the reduced density matrix is defined as $\rho^\psi_A:=tr_{\bar A}|\psi\rangle\langle \psi|$. One could obtain the spectral projection for $\rho^\psi_A$ by Eq.(\ref{Riesz_projection}), which will be denoted as $\mathbb{P}_\psi(\lambda_i)$ in the following. We would like to consider the projected state $\mathbb{P}_\psi(\lambda_i)|\psi\rangle$. The projected staet $\mathbb{P}_\psi(\lambda_i)|\psi\rangle$ is unnormalized. Its normalization is given by
\bea
\langle \psi| \mathbb{P}_\psi(\lambda_i)^2|\psi\rangle=\lambda_i tr \mathbb{P}_\psi(\lambda_i)=\lambda_i\mathcal{P}(\lambda_i).
\eea
  Let us define the normalized projection state
\bea\label{projection_state_norm}
|\phi\rangle_{\lambda_i}:= \frac{\mathbb{P}_\psi(\lambda_i)|\psi\rangle}{\sqrt{\langle \psi |\mathbb{P}_\psi(\lambda_i)|\psi\rangle}}.
\eea
Similarly, one could define the projector $\bar{\mathbb{P}}_\psi(\lambda_i)$ for the reduced density matrix $\rho_{\bar A}$. By Schmidt decomposition, we would expect $\mathbb{P}_\psi(\lambda_i)|\psi\rangle =\bar{\mathbb{P}}_\psi(\lambda_i)|\psi\rangle$.

For any operators $\mathcal{O}_A$ locted in subsystem $A$, the expectation value in the projected state is given by
\bea\label{expectation_value}
&&\langle \psi|\mathbb{P}_\psi(\lambda_i)\mathcal{O}_A\mathbb{P}_\psi(\lambda_i)|\psi\rangle= \lambda_i tr(\mathcal{O}_A \mathbb{P}_\psi(\lambda_i))\nn \\
&&\phantom{\langle \psi|\mathbb{P}_\psi(\lambda_i)\mathcal{O}_A\mathbb{P}_\psi(\lambda_i)|\psi\rangle}=\frac{1}{2\pi i} \int_{\gamma^i} dz tr(R_{\rho^\psi_A}(z)\mathcal{O}_A)\nn \\
&&\phantom{\langle \psi|\mathbb{P}_\psi(\lambda_i)\mathcal{O}_A\mathbb{P}_\psi(\lambda_i)|\psi\rangle}=\frac{1}{2\pi i} \int_{\gamma^i} dz \sum_{n=1}^\infty \frac{tr[(\rho^\psi_A)^n\mathcal{O}_A]}{z^n}.
\eea
The expectation value can be evaluated with knowing $tr[(\rho^\psi_A)^n\mathcal{O}_A]$, which can be obtained by using replica method in QFTs. It is also straightforward to show that
\bea
\langle \psi|\mathbb{P}_\psi(\lambda_i)\mathcal{O}_A\mathbb{P}_\psi(\lambda_j)|\psi\rangle=tr \left(\rho_A \mathbb{P}_\psi(\lambda_i)\mathcal{O}_A\mathbb{P}_\psi(\lambda_j)\right)=\lambda_i \delta_{\lambda_i,\lambda_j}tr(\mathcal{O}_A \mathbb{P}_\psi(\lambda_i)).
\eea
By using the Euclidean path integral and replica method we have $tr\rho_A^n=Z_n/Z_1^n$, where $Z_n$ is the partition function on the n-sheeted surface $\Sigma_n$. By the definition of the one-point function we have
\bea
tr(\rho_A^n \mathcal{O}_A)=\langle \mathcal{O}_A\rangle_{\Sigma_n} tr\rho_A^n.
\eea 
One could also define the expectation value in the normalized projected state
\bea\label{expectation_lambda}
\langle \mathcal{O}_A\rangle_{\lambda_i,\psi}:=~_{\lambda_i}\langle \phi|\mathcal{O}_A|\phi\rangle_{\lambda_i}.
\eea
In \cite{Guo:2020roc} the author defines the quantity
\bea\label{Expectation_alt}
\mathcal{P}_{\mathcal{O}_A}(\lambda):= \sum_{i} \langle i|\mathcal{O}_A |i\rangle \delta(\lambda_i-\lambda),
\eea
where $|i\rangle$ is the eigenstate with eigenvalue $\lambda_i$. $\mathcal{P}_{\mathcal{O}_A}(\lambda)$ is the density of eigenvalues for $\mathcal{O}_A=I$ where $I$ is the  identity operator. By similar arguments as in Section \ref{section_density}, we expect that $\mathcal{P}_{\mathcal{O}_A}(\lambda_i)=\langle \psi|\mathbb{P}_\psi(\lambda_i)\mathcal{O}_A\mathbb{P}_\psi(\lambda_i)|\psi\rangle$. Further, we can define 
\bea
\bar{\mathcal{P}}_{\mathcal{O}_A}(\lambda):= \frac{\mathcal{P}_{\mathcal{O}_A}(\lambda)}{\mathcal{P}(\lambda)}.
\eea
It is expected that
\bea\label{averaged}
\bar{\mathcal{P}}_{\mathcal{O}_A}(\lambda_i)=~_{\lambda_i}\langle \phi|\mathcal{O}_A|\phi\rangle_{\lambda_i}.
\eea
 As shown in \cite{Guo:2020roc} $\mathcal{P}_{\mathcal{O}_A}(\lambda)$ can be evaluated using the inverse Laplace transformation. In the following section, we will demonstrate that these two methods yield the same results for certain examples. The inverse Laplace transformation method is particularly useful when discussing states with holographic duals.

\subsection{Example}\label{section_example}
In this section, we will use a simple example to demonstrate how to evaluate $\mathcal{P}_\psi(\lambda_i)$ and $\mathcal{P}_{\mathcal{O}_A}(\lambda_i)$ by using Eq.(\ref{density}) and Eq.(\ref{expectation_value}). The simplest example is a single interval in the vacuum state of 2-dimensional CFTs. 
\subsubsection{Density of eigenstates}\label{section_Density_interval}
For one interval $A$ with length $L$ in vacuum state of 2-dimensional CFTs, the R\'enyi entropy is given by 
\bea\label{Renyivacuum}
S_n(\rho_A)=\frac{1+n}{n}b,
\eea
where $b=\frac{c}{6}\log\frac{L}{\epsilon}$ and $e^{-b}:=\lambda_m=\lim_{n\to \infty}S_n(\rho_A)$. According to (\ref{density}) we have
\bea\label{vacuum_density}
&&\lambda_i \mathcal{P}(\lambda_i)=\frac{1}{2\pi i} \int_{\gamma^i}dz \sum_{n=1}^\infty \frac{1}{(z e^b)^n}e^{b/n}\nn \\
&&\phantom{\lambda \mathcal{P}(\lambda_i)}=\frac{1}{2\pi i} \int_{\gamma^i}dz \sum_{k=0}^\infty \frac{b^k}{k!}Li_k(\frac{1}{ze^b}),
\eea
where $Li_k(x)$ is the polylogarithm function. We can express the polylogarithm function as the Bose-Einstein distribution 
\bea\label{bose_einstein}
Li_k(x)=\frac{1}{\Gamma(k)}\int_0^\infty dt \frac{t^{k-1}}{e^t/x -1}.
\eea
Eq.(\ref{vacuum_density}) can be expressed as
\bea
\lambda_i \mathcal{P}(\lambda_i)=\frac{1}{2\pi i} \int_0^\infty dt\int_{\gamma^i}dz \frac{e^{-b-t} \sqrt{b}I_1(2\sqrt{bt})}{\sqrt{t}(z-e^{-b-t})},
\eea
where $I_n(x)$ is the modified Bessel function of the first kind. The contour $\gamma^i$ is around the eigenvalue $\lambda_i$, thus the contour integration in the above formula would give a delta function, that is
\bea\label{density}
\lambda_i \mathcal{P}(\lambda_i)=\int_0^\infty dt \frac{e^{-b-t} \sqrt{b}I_1(2\sqrt{bt})}{\sqrt{t}}\delta(\lambda_i-e^{-b-t})= \frac{\sqrt{b}I_1(2\sqrt{-b(b+\log\lambda_i)})}{\sqrt{-b-\log\lambda_i}},
\eea
which is consistent with previous results \cite{Calabrese2008}\cite{Guo:2020roc}. However, it is important to note the disadvantage of the projection method. It seems difficult to obtain the density of eigenvalues at the maximal eigenvalues. As shown in \cite{Calabrese2008}, $\mathcal{P}(\lambda_m)$ would be a delta function $\delta (\lambda-\lambda_m)$.

\subsubsection{Expectation value of stress energy tensor in the projection state}\label{section_stress_vacuum}
We would like to use (\ref{expectation_value}) to evaluate the expectation value of stress energy $T(w)$ and $\bar T(\bar w)$  in the projected state with $|\psi\rangle=|0\rangle$ in 2-dimensional CFTs. The subsystem $A$ is an interval $(R_0,R_1)$ on $\tau=0$. By using the conformal transformation $\zeta=(\frac{w-R_0}{R_1-w})^{1/n}$,  the n-sheeted surface $\Sigma_n$ is mapped to complex $\zeta$-plane. With this one could obtain the expectation value of $T(w)$ on the n-sheeted surface $\Sigma_n$ 
\bea
\langle T(w)\rangle_{\Sigma_n}=\frac{c}{24}\frac{n^2-1}{n^2}\frac{(R_1-R_0)^2}{(w-R_0)^2(R_1-w)^2}.
\eea
Using (\ref{expectation_value}) we have
\bea
&&\langle 0|\mathbb{P}(\lambda_i)T(w)\mathbb{P}(\lambda_i)|0\rangle\nn \\
&&=\frac{1}{2\pi i} \int_{\gamma^i} dz \sum_{n=1}^\infty \frac{tr[(\rho^\psi_A)^n\mathcal{O}_A]}{z^n}\nn \\
&&=\frac{1}{2\pi i} \int_{\gamma^i} dz \sum_{n=1}^\infty \frac{tr[(\rho^\psi_A)^n] }{z^n}\langle T(w)\rangle_{\Sigma_n}\nn \\
&&=\frac{c}{24}\frac{(R_1-R_0)^2}{(w-R_0)^2(R_1-w)^2}\frac{1}{2\pi i} \int_{\gamma^i} dz \sum_{n=1}^\infty \frac{1}{(ze^b)^n}e^{b/n}\frac{n^2-1}{n^2}\nn\\
&&=\frac{c}{24}\frac{(R_1-R_0)^2}{(w-R_0)^2(R_1-w)^2}\frac{1}{2\pi i} \int_{\gamma^i} dz \sum_{k=0}^\infty \frac{b^k}{k!}\left( Li_{k+2}(\frac{1}{ze^b})-Li_{k}(\frac{1}{ze^b})\right).\nn
\eea
By using (\ref{bose_einstein}) we have
\bea\label{expectation_vacuum}
&&\langle 0|\mathbb{P}(\lambda_i)T(w)\mathbb{P}(\lambda_i)|0\rangle\nn \\
&&=\frac{c}{24}\frac{(R_1-R_0)^2}{(w-R_0)^2(R_1-w)^2}\frac{1}{2\pi i} \int_{\gamma^i} dz \int_0^\infty dt \frac{e^{-b-t}}{z-e^{-b-t}}\frac{(b-t)I_1(2\sqrt{bt})}{\sqrt{bt}}\nn \\
&&=\frac{c}{24}\frac{(R_1-R_0)^2}{(w-R_0)^2(R_1-w)^2}  \frac{\sqrt{b}I_1(2\sqrt{b(-b-\log\lambda_i)})}{\sqrt{-b-\log\lambda_i}}\left(1-\frac{-b-\log\lambda_i}{b} \right).\nn\\
~
\eea
By using the method in \cite{Guo:2020roc} one could obtain $\mathcal{P}_T(\lambda)$, which is consistent with the above result. Further we have the expectation value in the normalized state $|\Phi\rangle_{\lambda_i}$
\bea\label{expectation_vacuum_norm}
\langle T(w) \rangle_{\lambda_i,0}=\frac{c}{24}\frac{(R_1-R_0)^2}{(w-R_0)^2(R_1-w)^2} \left(1-\frac{-b-\log\lambda_i}{b} \right).
\eea
A similar formula can be derived for $\bar T(\bar w)$ by replacing $w$ by $\bar w$ in the expression. A notable fact is that the expectation value of $T(w)$ and $\bar T(\bar w)$ is divergent near the boundary of $A$. In the following we will demonstrate that this is a universal property for the projection states.

\subsubsection{Higher order correlator in the projection state}\label{section_higher_order}
Let us consider the two-point correlators of stress energy tensor in the projected state, that is  
\bea
\langle 0|\mathbb{P}(\lambda_i)T(w)T(w')\mathbb{P}(\lambda_i)|0\rangle,
\eea
where the operators are located in subsystem $A$. To calculate the correlator we should evaluate $\langle T(w)T(w')\rangle_{\Sigma_n}$. By using the conformal transformation $\zeta=(\frac{w-R_0}{R_1-w})^{1/n}$ we have
\bea\label{twopoint_sigma}
&&\langle T(w)T(w')\rangle_{\Sigma_n}\nn \\
&&=\left( \frac{d\zeta}{dw}\right)^2\left( \frac{d\zeta'}{dw'}\right)^2\langle T(\zeta)T(\zeta')\rangle+\left(\frac{c}{12}\right)^2 \{\zeta;w \}\{\zeta';w' \}\nn\\
&&=\frac{(R_1-R_0)^{4}}{(w-R_0)^2(R_1-w)^2(w'-R_0)^2(R_1-w')^2}\left[\frac{(\zeta \zeta')^{2}}{n^4}\frac{c/2}{(\zeta-\zeta')^4}+\left(\frac{c}{24}\frac{n^2-1}{n^2}\right)^2\right],\nn\\
~
\eea
where $\zeta'=(\frac{w'-R_0}{R_1-w'})^{1/n}$ and $\{z;w\}$ denotes the Schwarzian derivative. With this we have 
\bea\label{expectation_twopoint}
&&\langle 0|\mathbb{P}(\lambda_i)T(w)T(w')\mathbb{P}(\lambda_i)|0\rangle\nn \\
&&=\frac{1}{2\pi i} \int_{\gamma^i} dz \sum_{n=1}^\infty \frac{tr[(\rho^\psi_A)^nT(w)T(w')]}{z^n}\nn \\
&&=\frac{1}{2\pi i} \int_{\gamma^i} dz \sum_{n=1}^\infty \frac{tr[(\rho^\psi_A)^n] }{z^n}\langle T(w)T(w')\rangle_{\Sigma_n}.\nn 
\eea
Taking (\ref{twopoint_sigma}) into (\ref{expectation_twopoint}) one could obtain the two-point correlator of $T(w)$ in the projection state. Let us define $a:=\frac{(w-R_0)(R_1-w')}{(R_1-w)(w'-R_0)}$. We will consider the case $a\ll 1$, that is the distance between the two operators are small. In Appendix.\ref{Appendix_higher_correlator} we show the details of the calculations. Using the results in Appendix.\ref{Appendix_higher_correlator} we have
\bea
&&\langle 0|\mathbb{P}(\lambda_i)T(w)T(w')\mathbb{P}(\lambda_i)|0\rangle\nn \\
&&=\frac{c}{12}\frac{(R_1-R_0)^4}{a^2 (R_1-w)^4(w'-R_0)^4} \left[\frac{a(a^2+4a+1)}{(a-1)^4} \frac{\sqrt{b} I_1(2\sqrt{b t_i})}{\sqrt{t_i}}-\frac{a}{(a-1)^2}\frac{\sqrt{t_i} I_1(2\sqrt{bt_i})}{\sqrt{b}} \right]\nn \\
&&+\frac{c^2}{576}\frac{(R_1-R_0)^4}{a^2 (R_1-w)^4(w'-R_0)^4} \Big[ \frac{\sqrt{b} I_1\left(2 \sqrt{bt_i}\right)}{\sqrt{t_i}}-\frac{4 \sqrt{t_i} I_1\left(2 \sqrt{bt_i}\right)}{\sqrt{b}}+\frac{6 t_i^{3/2} I_3\left(2 \sqrt{bt_i}\right)}{b^{3/2}}\nn \\
&&\phantom{+\frac{c^2}{576}\frac{(R_1-R_0)^4}{a^2 (R_1-w)^4(w'-R_0)^4} \Big[}-\frac{4 t_i^{5/2} I_5\left(2 \sqrt{bt_i}\right)}{b^{5/2}}+\frac{t_i^{7/2} I_7\left(2 \sqrt{bt_i}\right)}{b^{7/2}}\Big]+O((a-1)^0).\nn\\
\eea
We can also obtain
\bea\label{two_point_stress_projected}
&&\langle T(w)T(w')\rangle_{\lambda_i,0}\nn \\
&&=\frac{c}{12}\frac{(R_1-R_0)^4}{a^2 (R_1-w)^4(w'-R_0)^4} \left[\frac{a(a^2+4a+1)}{(a-1)^4} -\frac{a}{(a-1)^2}\frac{t_i}{b} \right]\nn \\
&& +\frac{c^2}{576}\frac{(R_1-R_0)^4}{a^2 (R_1-w)^4(w'-R_0)^4} \Big[1-\frac{4 t_i}{b}+6 \left(\frac{t_i}{b}\right)^2\frac{ I_3}{I_1}-4 \left(\frac{t_i}{b}\right)^3\frac{ I_5}{I_1}+\left(\frac{t_i}{b}\right)^4\frac{ I_7}{I_1}\Big],\nn\\
~
\eea
where we use the shorthand notation $I_n=I_n(2\sqrt{bt})$. The expression of the two-point correlator is complicated. For the holographic theory we would have a large central  charge $c$. Since $b\sim O(c)$ we expect $b$ is also large.  In the limit $c\to +\infty$, we have $I_n(2\sqrt{b t})/I_1(2\sqrt{bt})\to 1$. The above expression would be
\bea\label{Projected_state_holography}
&&\langle T(w)T(w')\rangle_{\lambda_i,0} \nn \\
&&=\frac{c}{12}\frac{(R_1-R_0)^4}{a^2 (R_1-w)^4(w'-R_0)^4} \left[\frac{a(a^2+4a+1)}{(a-1)^4} -\frac{a}{(a-1)^2}\frac{t_i}{b} \right]\nn \\
&&+\frac{c^2}{576}\frac{(R_1-R_0)^4}{a^2 (R_1-w)^4(w'-R_0)^4} \left(1-\frac{t_i}{b} \right)^2+O((a-1)^0,c^{-1}).
\eea
An interesing fact is 
\bea\label{twopointcorrelator_T}
\langle T(w)T(w')\rangle_{\lambda_i,0}-\langle T(w)\rangle_{\lambda_i,0}\langle T(w')\rangle_{\lambda_i,0}\sim O(c).
\eea
We can define the operator $u(w):=\frac{T(w)}{\sqrt{c}}$, the expectation value of which is of order $O(\sqrt{c})$. Then we have
\bea
\lim_{c\to \infty}\left( \langle u(w)u(w')\rangle_{\lambda_i,0}-\langle u(w)\rangle_{\lambda_i,0}\langle u(w')\rangle_{\lambda_i,0}\right)=O(c^0),
\eea
which means the quantum fluctuation of the operator $u(w)$ is suppressed in the projected state. The opeartor $u(w)$ can be taken as a classical operator. 

By using the same method one could compute the higher order correlators of stress energy tensor , such as
\bea
\langle \prod_{i=1}^{m}T(w_i)\rangle_{\lambda_i,0},
\eea
where $m$ is an integer.
It is straightforward but tedious to show that the higher order connected correlators of $u(w)$  is suppressed in the semi-classical limit $c\to \infty$. 

\subsection{Projector and insertion of local operator}
From (\ref{expectation_vacuum}) and (\ref{expectation_vacuum_norm}) we find the expectation value of $T(w)$ is divergent near the boundary $R_0$ and $R_1$, which is very similar as the twist operator. It appears that the projector $\mathbb{P}(\lambda_i)$ only effects the boundary of the subsystem $A$. In the following, we will illustrate the difference that arises from the insertion of a local operator at the boundary of $A$.

The R\'enyi entropy can be computed by using twist operators formalism, $tr\rho_A^n$ can be expressed as correlator of twist oprators located at the boundary of $A$ in the  $n$-copied theories. An evidence for this is that the correlator involving of $T(w)$ and twist operators gives the correct result of the expectation value of $T(w)$ on the Riemann surface $\Sigma_n$. Assume $A$ is an interval $[R_0,R_1]$, for the vacuum state we have
\bea\label{twist_stressenergy}
\langle T(w)\rangle_{\Sigma_n}= \frac{\langle T(w) \sigma_n(R_0,R_0) \tilde{\sigma}_n(R_1,R_1)\rangle_{n}}{\langle \sigma_n(R_0,R_0) \tilde{\sigma}_n(R_1,R_1)\rangle_{n}},
\eea
where the subscript $n$ of the correlator denotes the $n$-copied states. We find that our result (\ref{expectation_vacuum_norm}) shows similar properties as the twist operators. Supposed that the local operators $\Pi(w,\bar w)$ with conformal dimension $h_{\Pi}=\bar h_{\Pi}$ are inserted at the boundary of $A$ in the vacuum state. By conformal Ward identity we would have the correlator
\bea
\langle T(w)\Pi(R_0,R_0)\Pi(R_1,R_1)\rangle=\frac{h_{\Pi}(R_1-R_0)^2}{(w-R_0)^2(R_1-w)^2}\langle\Pi(R_0,R_0)\Pi(R_1,R_1)\rangle.
\eea
Comparing with (\ref{expectation_vacuum}) the projected state $\mathbb{P}(\lambda_i)|0\rangle$  show the same divergent behavior as the insertion of local operators at the boundaries of $A$. If taking $h_\Pi=\frac{c}{24}(1-\frac{-b-\log\lambda_i}{b})$ one would have
\bea
\langle T(w)\rangle_{\lambda_i,0}=\frac{\langle T(w)\Pi(R_0,R_0)\Pi(R_1,R_1)\rangle}{\langle\Pi(R_0,R_0)\Pi(R_1,R_1)\rangle},
\eea
which is very similar as (\ref{twist_stressenergy}).
But here the correlator 
\bea
\langle\Pi(R_0,R_0)\Pi(R_1,R_1)\rangle=\frac{1}{(R_1-R_0)^{4 h_\Pi}}\ne \langle 0|\mathbb{P}(\lambda_i)|0\rangle=\frac{\sqrt{b}I_1(2\sqrt{b(-b-\log\lambda_i)})}{\sqrt{-b-\log\lambda_i}}.
\eea
From (\ref{expectation_vacuum}) we see the projection state cannot be explained as insertion  of local operators $\Pi$ at the boundary of $A$. This is expected since the projector should effect the whole subregion $A$ not just its boundaries.

The Eq.(\ref{expectation_vacuum_norm}) provides us a way to understand the projection state
\bea\label{projected_state_vacuum}
|0\rangle_{\lambda_i}:=\frac{\mathbb{P}(\lambda_i)|0\rangle \langle  0| \mathbb{P}(\lambda_i)}{\langle 0|\mathbb{P}(\lambda_i)|0\rangle}.
\eea
By a conformal transformation 
\bea\label{conformal}
\zeta=\left( \frac{w-R_0}{R_1-w}\right)^{\alpha},\quad \bar \zeta=\left( \frac{\bar w-R_0}{R_1-\bar w}\right)^{\alpha},
\eea
we find
\bea
&&\langle T(\zeta)\rangle=\left(\frac{d\zeta}{dw}\right)^{-2} \left[\langle T(w)\rangle_{\lambda_i,0}-\frac{c}{12}\{\zeta;w \} \right]\nn \\
&&\phantom{\langle T(\zeta)\rangle}=\left(\frac{dz}{dw}\right)^{-2} \frac{c}{24}\frac{(R_1-R_0)^2}{(w-R_0)^2(R_1-w)^2} \left(\alpha^2-\frac{-b-\log\lambda_i}{b} \right).
\eea
 If $\alpha=\sqrt{\frac{t_i}{b}}$ with $t_i:=-b-\log \lambda_i$, we would have $\langle T(\zeta)\rangle=0$. It appears the projection state (\ref{projected_state_vacuum}) is mapped to the vacuum state on the complex $z$ plane by the conformal transformation (\ref{conformal}). But we will show it is not exact correct. 

Let us consider the two point correlator of $T(w)$. If the projection state is conformally equal to the vacuum state on the $\zeta$ plane by the transformation (\ref{conformal}). We would have the two point correlator
\bea\label{two_point_defect}
&&\langle T(w)T(w')\rangle= \frac{\alpha^2(R_1-R_0)^4}{a^2 (R_1-w)^4(w'-R_0)^4}\frac{c/2}{\left(a^{\frac{\alpha}{2}}-a^{-\frac{\alpha}{2}}\right)}\nn \\
&&\phantom{\langle T(w)T(w')\rangle=}+\frac{c^2}{576}\frac{(R_1-R_0)^4}{a^2 (R_1-w)^4(w'-R_0)^4} \left(1-\alpha^2 \right)^2,
\eea
where $a:=\frac{(w-R_0)(R_1-w')}{(R_1-w)(w'-R_0)}$. One could expand the above result near $a=1$, which is different with the two point correlator in the projection state (\ref{two_point_stress_projected}). 

However, if we consider the holographic theory with large $c$, the two point function of $T(w)$ in the projection state is given by (\ref{Projected_state_holography}), which matches (\ref{two_point_defect}) at the leading order of  $O(c^2)$. 
By using projection states, one can construct new states through superposition. In the following Section \ref{section_geometry}, we will discuss the possible metric duals to these states. The conformal map will be helpful in understanding the gravity dual.

\subsection{Universal behavior for the projection state}\label{section_universal}

In this section we will only focus on 2-dimensional CFTs. Now let us consider the projection state $\mathbb{P}_\psi(\lambda_i)|\psi\rangle$ for the general pure state $|\psi\rangle$.  We would like to discuss the universal behavior of expectation value of $T(w)$  in the projection state. Assume one of the boundary of subsystem $A$ is located at $x'$. By using the twist operator formalism we have
\bea
tr[(\rho_A^\psi)^n T(w)]=\langle\Psi| T(w)\sigma_n(x')...|\Psi\rangle, 
\eea 
where $\sigma_n$ is the twist operator with conformal dimension $h_n=\frac{c}{24}(n-\frac{1}{n})$, $|\Psi\rangle:=|\psi\rangle_1\otimes...|\psi\rangle_i\otimes...\otimes|\psi\rangle_n$ is the state on $n$ copied CFT. We will consider the behavior of the result near the boundary of $A$, that is taking the limit $w\to x'$. Using  Ward identity we have
\bea\label{stress_twist}
&&\langle \Psi| T(w)\sigma_n(x')...|\Psi\rangle\nn \\
&&=\frac{h_n}{n(w-x')^2} \langle \Psi|\sigma_n(x')...|\Psi\rangle+\frac{1}{2n(w-x')}\partial_{x'}\langle \Psi|\sigma_n(x')...|\Psi\rangle+...\nn \\
&&=\frac{h_n}{n(w-x')^2} tr(\rho_A^\psi)^n+\frac{1}{2n(w-x')}\partial_{x'}tr(\rho_A^\psi)^n +...\; ,\nn
\eea  
where $...$ denotes the non-divergent terms. Note that there is a factor of $\frac{1}{2}$ in the coefficient of the divergent term of order $O((w-x')^{-1})$, which follows the fact that $\partial_w=\frac{1}{2}\p_x+ \frac{1}{2i}\p_\tau$.
According to Eq.(\ref{expectation_value}) we have
\bea
&&\langle \psi|\mathbb{P}_\psi(\lambda_i)T(w)\mathbb{P}_\psi(\lambda_i)|\psi\rangle\nn \\
&&=\frac{1}{2\pi i} \int_{\gamma^i} dz \sum_{n=1}^\infty \frac{h_n}{nz^n}\frac{1}{(w-x')^2} tr(\rho_A^\psi)^n+\frac{1}{2\pi i} \int_{\gamma^i} dz \sum_{n=1}^\infty \frac{1}{2nz^n}\frac{1}{(w-x')}\partial_{x'}tr(\rho_A^\psi)^n +...\nn \\
&&=\frac{1}{2\pi i} \int_{\gamma^i} dz \sum_{n=1}^\infty \sum_{i'} \frac{h_n \lambda_{i'}^n}{nz^n}\frac{1}{(w-x')^2}  +\frac{1}{2\pi i} \int_{\gamma^i} dz \sum_{n=1}^\infty \sum_{i'}\frac{\lambda_{i'}^{n-1}\partial_{x'}\lambda_{i'}}{2z^n}\frac{1}{(w-x')} +...\nn \\
&&=\frac{c}{24(w-x')^2}\sum_{i'} \frac{1}{2\pi i}  \int_{\gamma^i} dz  \left(\frac{\lambda_{i'}}{z-\lambda_{i'}}-Li_2(\frac{\lambda_{i'}}{z})\right)+\frac{1}{2(w-x')}\sum_{i'} \frac{1}{2\pi i}  \int_{\gamma^i} dz \frac{1}{z-\lambda_{i'}}\partial_{x'} \lambda_{i'}+...\nn\\
&&=\frac{c}{24(w-x')^2}\sum_{i'} \frac{1}{2\pi i}  \int_{\gamma^i} dz  \frac{\lambda_{i'}}{z-\lambda_{i'}}-\frac{c}{24(w-x')^2}\sum_{i'} \frac{1}{2\pi i}  \int_{\gamma^i} dz  \int_0^\infty dt \frac{\lambda_{i'}t e^{-t}}{z-\lambda_{i'}e^{-t}}\nn \\
&&+\frac{1}{2(w-x')}\sum_{i'} \frac{1}{2\pi i}  \int_{\gamma^i} dz \frac{1}{z-\lambda_{i'}}\partial_{x'} \lambda_{i'}+...,
\eea
where in the last step we use Eq.(\ref{bose_einstein}).
By using the fact that $\sum_{i'}\int_{\gamma^i}\frac{1}{z-\lambda_{i'}}=\sum_{i'}\delta_{\lambda_i,\lambda_{i'}}=\mathcal{P}(\lambda_i)$, we have
\bea
\sum_{i'} \frac{1}{2\pi i}  \int_{\gamma^i} dz  \frac{\lambda_{i'}}{z-\lambda_{i'}}=\lambda_i \mathcal{P}(\lambda_i).
\eea
We can also understand the above result as follows,
\bea
&&\sum_{i'} \frac{1}{2\pi i}  \int_{\gamma^i} dz  \frac{\lambda_{i'}}{z-\lambda_{i'}}=\frac{1}{2\pi i}  \int_{\gamma^i} dz \int_0^{\lambda_m}d\lambda \frac{\lambda}{z-\lambda}\mathcal{P}(\lambda)\nn \\
&&\phantom{\sum_{i'} \frac{1}{2\pi i}  \int_{\gamma^i} dz  \frac{\lambda_{i'}}{z-\lambda_{i'}}}=\int_0^{\lambda_m}d\lambda \lambda \mathcal{P}(\lambda) \delta(\lambda-\lambda_i)=\lambda_i \mathcal{P}(\lambda_i).
\eea
Similarly, we have
\bea
\sum_{i'} \frac{1}{2\pi i}  \int_{\gamma^i} dz \frac{1}{z-\lambda_{i'}}\partial_{x'} \lambda_{i'}=\frac{1}{2\pi i}  \int_{\gamma^i} dz \frac{1}{z-\lambda}\mathcal{P}_{x'}(\lambda),
\eea
where $\mathcal{P}_{x'}(\lambda):=\sum_i \frac{\p \lambda_i}{\p x'}\delta(\lambda_i-\lambda)$. $\mathcal{P}_{x'}(\lambda)$ is the function that is defined in \cite{Guo:2023tys}, which reflects the parameter dependence of the eigenvalues. We can work out 
\bea
\sum_{i'} \frac{1}{2\pi i}  \int_{\gamma^i} dz \frac{1}{z-\lambda_{i'}}\partial_{x'} \lambda_{i'}=\mathcal{P}_{x'}(\lambda_i).
\eea
 Lastly, the integration in the above formula can be worked out as follows,
\bea
&&-\sum_{i'} \frac{1}{2\pi i}  \int_{\gamma^i} dz  \int_0^\infty dt \frac{\lambda_{i'}t e^{-t}}{z-\lambda_{i'}e^{-t}}\nn \\
&&=-\sum_{i'}\int_0^{\infty}dt \delta(\lambda_i-\lambda_{i'}e^{-t})\lambda_{i'}t e^{-t} H(\lambda_{i'}-\lambda_i)\nn \\
&&=\sum_{i'}\log \frac{\lambda_i}{\lambda_{i'}}H(\lambda_{i'}-\lambda_i),
\eea
which can be written as the integration according to Eq.(\ref{continuous_integral})
\bea
\int_{\lambda_{i}}^{\lambda_m}d\lambda \log\frac{\lambda_i}{\lambda}\mathcal{P}(\lambda).
\eea
In summary, we have
\bea\label{universal_relation}
&&\langle \psi|\mathbb{P}_\psi(\lambda_i)T(w)\mathbb{P}_\psi(\lambda_i)|\psi\rangle\nn \\
&&=\frac{c}{24(w-x')^2} \left(\lambda_i \mathcal{P}(\lambda_i)+\int_{\lambda_{i}}^{\lambda_m}d\lambda \log\frac{\lambda_i}{\lambda}\mathcal{P}(\lambda) \right)+\frac{1}{2(w-x')} \mathcal{P}_{x'}(\lambda_i)+...\; .\nn\\~
\eea
One could also obtain the expectation value in the normalized projection state $|\Phi\rangle_{\lambda_i}$
\bea\label{universal_relation_norm}
&&\langle T(w)\rangle_{\lambda_i,\psi}\nn \\
&&=\frac{c}{24(w-x')^2} \left(1 +\frac{1}{\lambda_i}\int_{\lambda_{i}}^{\lambda_m}d\lambda \frac{\mathcal{P}(\lambda)}{\mathcal{P}(\lambda_i)}\log\frac{\lambda_i}{\lambda} \right)+\frac{1}{2(w-x')\lambda_i} \frac{\mathcal{P}_{x'}(\lambda_i)}{\mathcal{P}(\lambda_i)}+...\; .\nn\\~
\eea
The above result shows the expectation value of $T(w)$ in the projection state $\mathbb{P}(\lambda_i)|\psi\rangle$ is divergent near the boundary of the subsystem $A$. The leading divergent term is of order $O((w-x')^{-2})$, where $x'$ denotes the coordinate of the boundary. The coefficient of this term only depends on the eigenvalue $\lambda_i$ and density of eigenvalue $\mathcal{P}(\lambda_i)$. The next leading divergent term is of order $O((w-x')^{-1})$. The coefficient is only dependent with the function $\mathcal{P}_{x'}(\lambda)$ \cite{Guo:2023tys}. 

\subsection{Inverse Laplace method}\label{section_universal_ILM}

We can also evlauate the universal divergent terms of expectation value of $T(w)$ in the projection state by inverse Laplace transformation method, which is more straightforward \cite{Guo:2020roc}. We would like to study the expectation value of the stress energy tensor $T(w)$ in the projection state. Acoording to Eq.(\ref{Expectation_alt}) equally we can consider
\bea
\mathcal{P}_{T(w)}(\lambda):= \sum_{i} \langle i| T(w)|i\rangle \delta(\lambda_i-\lambda),
\eea
which can be taken as the summation of expectation value in the eigenstates with eigenvalue $\lambda$.
By definition we would have
\bea
tr(\rho_A^n T(w))=\sum_i \lambda_i^n \langle i| T(x)|i\rangle=\int_0^{\lambda_m}d\lambda \lambda^n \mathcal{P}_{T(w)}(\lambda),
\eea
where $\lambda_m$ is the maximal eigenvalues of $\rho_A^\psi$. Define the variable $\lambda=e^{-b-t}$, where $b=\lim_{n\to\infty} S_n(\rho_A^\psi)$. Thus we have
\bea
\int_0^{\infty}dt e^{-b-t}\mathcal{P}_{T(w)}(\lambda)=e^{n b}tr(\rho_A^n T(w)).
\eea
By inverse Laplace transformation we have
\bea
\mathcal{P}_{T(w)}(\lambda)=e^{b+t}
\mathcal{L}^{-1}[ e^{n b}tr(\rho_A^n T(w))](t).
\eea
On the other hand $tr(\rho_A^n T(w))$ can be obtained by using twist operator formulism. More precisely, we have
\bea
tr(\rho_A^n T(w))=\langle \Psi| T(w)\sigma_n(x')...|\Psi\rangle,
\eea
where $...$ denotes other twist operators,  $|\Psi\rangle:=|\psi\rangle_0...|\psi\rangle_i...|\psi\rangle_n$ is the state on $n$ copied CFT. According to (\ref{stress_twist}) we can directly obtain
\bea
&&\mathcal{P}_{T(w)}(\lambda)=\frac{c}{24(w-x')^2}\sum_{i'}\left(\lambda^{-1} \delta(\log\frac{\lambda_{i'}}{\lambda})+\lambda H(\log\frac{\lambda_i}{\lambda})\log\frac{\lambda_{i'}}{\lambda} \right)+\frac{1}{2(w-x')}\lambda^{-1}\delta(\log\frac{\lambda_{i'}}{\lambda})\nn\\
&&\phantom{\mathcal{P}_{T(w)}(\lambda)}=\frac{c}{24(w-x')^2}\int_0^{\lambda_m}d\lambda'\left(\lambda^{-1} \delta(\log\frac{\lambda'}{\lambda})\mathcal{P}(\lambda')+\lambda^{-1} H(\log\frac{\lambda'}{\lambda})\log\frac{\lambda'}{\lambda} \mathcal{P}(\lambda')\right)\nn \\
&&\phantom{\mathcal{P}_{T(w)}(\lambda)=}+\frac{1}{2(w-x')}\int_0^{\lambda_m}d\lambda' \frac{\lambda}{\lambda'}\delta(\log\frac{\lambda'}{\lambda})\mathcal{P}(\lambda')\mathcal{P}_{x'}(\lambda')\nn \\
&&\phantom{\mathcal{P}_{T(w)}(\lambda)}=\frac{c}{24(w-x')^2}\left(\mathcal{P}(\lambda)+\frac{1}{\lambda} \int_{\lambda}^{\lambda_m}d\lambda' \log\frac{\lambda'}{\lambda} \mathcal{P}(\lambda') \right)+\frac{1}{2(w-x')}\frac{\mathcal{P}(\lambda)\mathcal{P}_{x'}(\lambda)}{\lambda},
\eea
where in the second step we use Eq.(\ref{continuous_integral}). We can calculate $\bar{\mathcal{P}}_{T(w)}(\lambda)$ by using the definition (\ref{averaged}). It is obvious that for the divergent terms 
$\bar{\mathcal{P}}_{T(w)}(\lambda_i)=\langle T(w)\rangle_{\lambda_i,\psi}$.

\subsection{Check for the universal relation}

We can decompose $\langle \psi|T(w)|\psi\rangle$ as
\bea
\langle \psi|T(w)|\psi\rangle=\sum_i \langle \psi|\mathbb{P}_\psi(\lambda_i)T(w)\mathbb{P}_\psi(\lambda_i)|\psi\rangle.
\eea
Assume the expectation value $\langle \psi|T(w)|\psi\rangle$ is smooth near the boundary of $A$.  Thus we would have the constraints
\bea
&&\sum_i\left(\lambda_i \mathcal{P}(\lambda_i)+\int_{\lambda_{i}}^{\lambda_m}d\lambda \log\frac{\lambda_i}{\lambda}\mathcal{P}(\lambda) \right)=0, \label{constraint}\\
&&\sum_i\mathcal{P}_{x'}(\lambda_i)=0.
\eea
It can be shown that 
\bea
\sum_i\mathcal{P}_{x'}(\lambda_i)=\sum_i \partial_{x'}\lambda_i=0,
\eea
which follows the fact $\sum_i\lambda_i=1$. We can write (\ref{constraint}) as integral and show
\bea
\int_{0}^{\lambda_m}d\lambda\lambda \mathcal{P}(\lambda)+\int_0^{\lambda_m}d\lambda'\int_{\lambda'}^{\lambda_m}d\lambda \log\frac{\lambda'}{\lambda}\mathcal{P}(\lambda) =0.
\eea

In Section \ref{section_stress_vacuum} we evaluate the expectation value of stress energy tensor in the projected state $\mathbb{P}(\lambda_i)|0\rangle$.  The result (\ref{expectation_vacuum_norm}) shows $\langle T(w)\rangle_{\lambda_i,0}$ is divergent near the boundary $R_0$. We can expand the result (\ref{expectation_vacuum_norm}) near $R_0$ as
\bea\label{expand_expectation}
\langle T(w)\rangle_{\lambda_i,0}=\left[\frac{c}{24 (w-R_0)^2}+\frac{c}{12 (R_1-R_0) (w-R_0)}\right]\left(1-\frac{-b-\log\lambda_i}{b} \right)+...\; .
\eea
The leading divergent term is of order $O((w-R_0)^{-2})$, next leading divergent term is of order $O((w-R_0)^{-1})$, which is consistent with the general result (\ref{universal_relation_norm}). Now let check the coefficients of the divergent terms. By using (\ref{density}) we have\footnote{In the calculation we should include the  contribution from the delta function of maximal eigenvalue.}
\bea
\int_{\lambda_{i}}^{\lambda_m}d\lambda \frac{\mathcal{P}(\lambda)}{\mathcal{P}(\lambda_i)}\log\frac{\lambda_i}{\lambda}=\int_0^{t_i}dt e^{-b-t}\frac{\mathcal{P}(e^{-b-t})}{\mathcal{P}(e^{-b-t_i})}(t-t_i)=-\frac{t_i}{b}\lambda_i,
\eea
where we define $\lambda=e^{-b-t}$ and $\lambda_i=e^{-b-t_i}$. Taking the above result into (\ref{universal_relation_norm}), we find the coefficient of order $O((w-R_0)^{-2})$ is consistent with (\ref{expand_expectation}).

In \cite{Guo:2023tys} we calculate the function $\mathcal{P}_L(\lambda)$, where $L:=R_1-R_0$ is the length  of the interval $A$. By using the notation in this paper for $\lambda_i \ne \lambda_m$ the result is
\bea
\mathcal{P}_L(\lambda_i)=-\frac{c}{6 (R_1-R_0)}\lambda_i  (1-\frac{t_i}{b})\mathcal{P}(\lambda_i).
\eea

By translation invariance one can show the eigenvalues only depend on the length $L$, thus we have $\mathcal{P}_{R_0}(\lambda):=\sum_i \frac{\p \lambda_i}{\p R_0}\delta(\lambda_i-\lambda)=-\mathcal{P}_{L}(\lambda)$. According to (\ref{universal_relation_norm}) the coefficient of the order of $O((w-R_0)^{-1})$ is given by 
\bea
\frac{c}{12(R_1-R_0)} (1-\frac{t_i}{b}),
\eea
which is consistent with (\ref{expand_expectation}).

\section{Projection state and fixed area state}\label{section_geometry}
In this section, we discuss the holographic aspect of the projection state $\mathbb{P}(\lambda_i)|\psi\rangle$. By definition, it is evident that the reduced density matrix of $A$ for the projection state 
would have flat spectrum. Consequently, the R\'enyi entropy is independent of $n$. In \cite{Akers:2018fow} and \cite{Dong:2018seb}, the authors construct the so-called fixed area state using gravitational path integrals and quantum error correction codes. For further studies on this topic, see \cite{Dong:2019piw}-\cite{Dong:2022ilf}. The fixed area state also has a flat spectrum. In \cite{Guo:2021tzs}, we propose a candidate for the fixed area state in dual CFTs using the projection states. Furthermore, one can construct the area operator $\hat A$, whose expectation value gives the area of the minimal surface in the bulk. Thus, the Ryu-Takayanagi formula can be written as the expectation value of the area operator. In this section, we further study the relationship between the projection state and the fixed area state. We show that the projection state with fixed $\lambda_i$ cannot be dual to a bulk geometry. Instead, by suitably superposing the projection states $\mathbb{P}(\lambda_i)|\psi\rangle$, one can obtain the correct state. We conclude this by comparing the partition function and R\'enyi entropy on both the gravity and field theory sides. Additionally, we demonstrate consistency with the results in \cite{Guo:2021tzs}, clarifying some subtleties.

\subsection{Quantum fluctuations in the projection state}\label{Fluctuations_projection}

In Section \ref{section_stress_vacuum} we calculate the expectation value of $T(w)$ in the projected state $\mathbb{P}(\lambda_i)|0\rangle$. In AdS$_3$ the bulk metric is determined by the expectation value of $T(w)$ and $\bar T(\bar w)$. We can use the expectation value of stress energy tensor to construct the metric of bulk geometry. 
In the Appendix.\ref{section_appendix_metric} we review how to construct the bulk geometry in AdS$_3$. 

Firstly, let us study the general features of projection states. Generally, the field theory dual to gravity is expected to be a gapped large-$N$ CFT\cite{Aharony:1999ti}\cite{Harlow:2018fse}. The bulk Newton constant $G\sim \frac{1}{N^2}$. In the context of AdS$_3$/CFT$_2$, we have the well-konwn Brown-Henneaux relation $c=\frac{3}{2G}$. Thus the semi-classical limit $G\to 0$ corresponds the large $N$ limit $N\to \infty$. If a given state $|\Psi\rangle$ can effectively be described by a bulk geometry, it is expected the quantum fluctuation should be small in the semi-classical limit. For single trace operators $\mathcal{O}_i$, the connected two-point correlator should satisfy
\bea\label{correlator_two_point}
\langle \Psi| \mathcal{O}_i \mathcal{O}_j |\Psi\rangle - \langle \Psi| \mathcal{O}_i|\Psi\rangle \langle \Psi| \mathcal{O}_j|\Psi\rangle 
\sim O(N^2).
\eea
We can also consider the rescaled operator $O_i:=\mathcal{O}_i/N$, the connected $k$-point correlators should be of order $O(N^{2-k})$\cite{Harlow:2018fse}\cite{Witten:2021unn}. For example, let us consider the stress energy tensor $T(w)$, in the vacuum state or thermal state we would have $tr T(w)\sim c \sim N^2$. One could define $u(w):= T(w)/\sqrt{c}$. It can be shown that $\langle u(w)u(w')\rangle-\langle u(w)\rangle \langle u(w')\rangle\sim O(c^0)$ in the vacuum state or thermal state. 

In Sections \ref{section_stress_vacuum} and \ref{section_higher_order}, we study the correlation functions of $T(w)$ in the projection state $\mathbb{P}(\lambda_i)|0\rangle$. The result (\ref{twopointcorrelator_T}) satisfies the required scaling behavior for a state with a holographic dual. Higher-point correlators also exhibit the same scaling behavior as those in the vacuum or thermal state. This implies that the quantum fluctuations of the operators in the projection state $\mathbb{P}(\lambda_i)|0\rangle$ is suppressed in the large $N$ limit. Thus it seems the projection state $\mathbb{P}(\lambda_i)|0\rangle$ may have a well-defined bulk geometry.  However, one should bear in mind that condition (\ref{correlator_two_point}) is only a basic requirement for a state to have a bulk geometric dual. The physical quantities must also be consistent when computed from both the bulk and the field theory sides. In the following, we will demonstrate that the state $\mathbb{P}(\lambda_i)|0\rangle$ cannot be dual to a bulk geometry, as it fails to produce the correct physical quantities as computed from the given bulk geometry.

Now let us consider more general case. Assume one has a state $|\Psi\rangle$ which is dual to a bulk geometry. Thus it satisfies the condition (\ref{correlator_two_point}).  One could construct the projection state $\mathbb{P}_{\Psi}(\lambda_i)|\Psi\rangle$. We would like to study the connected correction functions in the projection state. To do this we need to evaluate the two-point correlator
\bea
\langle \mathcal{O}_i \mathcal{O}_j\rangle_{\lambda_i,\Psi}:= \frac{\langle \Psi|\mathbb{P}_{\Psi}(\lambda_i) \mathcal{O}_i \mathcal{O}_j \mathbb{P}_{\Psi}(\lambda_i)|\Psi\rangle}{\langle \Psi|\mathbb{P}_{\Psi}(\lambda_i)|\Psi\rangle}.
\eea
Let us first consider the density of eigenvalues $\mathcal{P}_\Psi(\lambda_i)=\langle \Psi|\mathbb{P}_{\Psi}(\lambda_i)|\Psi\rangle$. If $|\Psi\rangle$ is a geometry state, it can be shown that $\mathcal{P}_\Psi(\lambda_i)$ can be evaluated by using saddle point approximation \cite{Guo:2021tzs}. We could calculate $\mathcal{P}_\Psi(\lambda_i)$ by using inverse Laplace transformation
\bea
&&\lambda_i\mathcal{P}_{\Psi}(\lambda_i)= \mathcal{L}^{-1}\left[e^{nb^\Psi}tr(\rho^\Psi_A)^n  \right]\nn \\
&&\phantom{\lambda_i\mathcal{P}_{\Psi}(\lambda_i)}=\frac{1}{2\pi i} \int_{\gamma-i T}^{\gamma+i T}dn e^{nt_i+n b^\Psi+(1-n)S_n(\rho_A^\Psi)},
\eea
where $t_i=-\log \lambda_i$ and $S_n(\rho_A^\Psi)$ is the R\'enyi entropy, $b^\Psi=\lim_{n\to \infty}S_n(\rho_A^\Psi)$, $\gamma$ is chosen for the convergence of the integration. According to the holographic R\'enyi entropy formula, we expect $S_n\sim b^\Psi\sim O(1/G)$.
Define the function
\bea\label{exp}
s_n:= nt_i+n b^\Psi+(1-n)S_n(\rho_A^\Psi).
\eea
 In the semiclassical limit $G\to 0$, the integration can be evaluated by saddle point approximation, that is to solve the equation 
\bea
\partial_n s_n=t_i+ b^\Psi+\partial_n[(1-n)S_n(\rho_A^\Psi)]=0.
\eea
Assume the solution of the above equation is $n^*(\lambda_i)$. With further using holographic R\'enyi entropy  formula \cite{Dong:2016fnf}, see also the modification \cite{Dong:2023bfy}, we can express the density of eigenvalues as follows,
\bea\label{lambdaPlambda_geometry}
\lambda_i \mathcal{P}_\Psi(\lambda_i)\simeq e^{\frac{\mathcal{B}_{n^*}}{4G}},
\eea
where $\mathcal{B}_{n^*}$ denotes the area of the cosmic brane tension $\mu_{n^*}=\frac{n^*-1}{4Gn^*}$, see \cite{Guo:2021tzs} for details. 
The two-point correlator $\langle \Psi|\mathbb{P}_{\Psi}(\lambda_i) \mathcal{O}_i \mathcal{O}_j \mathbb{P}_{\Psi}(\lambda_i)|\Psi\rangle$ is equal to consider the function
\bea
\mathcal{P}_{\mathcal{O}_i \mathcal{O}_j}(\lambda_i)=\sum_\alpha \delta(\lambda_i-\lambda_{i_\alpha})\langle i_\alpha|  \mathcal{O}_i \mathcal{O}_j |i_\alpha\rangle,
\eea
which can be computed by inverse Laplace transformation. Define $\lambda_i=e^{-b-t_i}$ we have
\bea
&&\lambda_i\mathcal{P}_{\mathcal{O}_i \mathcal{O}_j}(\lambda_i)= \mathcal{L}^{-1}\left[e^{nb^\Psi}tr(\rho^\Psi_A)^n \langle \mathcal{O}_i \mathcal{O}_j \rangle_{\Sigma_n,\Psi} \right]\nn \\
&&\phantom{\lambda_i\mathcal{P}_{\mathcal{O}_i \mathcal{O}_j}(\lambda_i)}=\frac{1}{2\pi i} \int_{\gamma-i T}^{\gamma+i T}dn e^{s_n}\langle \mathcal{O}_i \mathcal{O}_j \rangle_{\Sigma_n,\Psi},
\eea
where $s_n$ is given by (\ref{exp}), $\langle \mathcal{O}_i \mathcal{O}_j \rangle_{\Sigma_n,\Psi}$ denotes the two-point correlators on the n-sheeted manifold $\Sigma_n$. Since $\mathcal{O}_i$ are single trace operators, the two-point correlators on $\Sigma_n$ should scale as $O(N^m)$, where $m$ is some positive constant. The above integration can also be evaluated by using saddle point approximation with solving 
\bea
\partial_n \left( s_n +\log \langle \mathcal{O}_i \mathcal{O}_j \rangle_{\Sigma_n,\Psi} \right)=0.
\eea
In the semiclassical limit or large $N$ limit the solution of the above equation is still given by $n_*(\lambda_i)$. Thus we have
\bea
\lambda_i\mathcal{P}_{\mathcal{O}_i \mathcal{O}_j}(\lambda_i)=e^{\frac{\mathcal{B}_{n^*}}{4G}} \langle \mathcal{O}_i \mathcal{O}_j \rangle_{\Sigma_{n^*},\Psi}.
\eea
Thus we have 
\bea
\frac{\langle \Psi|\mathbb{P}_{\Psi}(\lambda_i) \mathcal{O}_i \mathcal{O}_j \mathbb{P}_{\Psi}(\lambda_i)|\Psi\rangle}{\langle \Psi|\mathbb{P}_{\Psi}(\lambda_i)|\Psi\rangle}=\langle \mathcal{O}_i \mathcal{O}_j \rangle_{\Sigma_{n^*},\Psi}.
\eea
Similarly, we can evaluate the one-point function
\bea
\frac{\langle \Psi|\mathbb{P}_{\Psi}(\lambda_i) \mathcal{O}_i\mathbb{P}_{\Psi}(\lambda_i)|\Psi\rangle}{\langle \Psi|\mathbb{P}_{\Psi}(\lambda_i)|\Psi\rangle}=\langle \mathcal{O}_i  \rangle_{\Sigma_{n^*},\Psi}
\eea
Thus the connected two-point correlator is 
\bea\label{Projection_factor}
\langle \mathcal{O}_i \mathcal{O}_j \rangle_{\lambda_i,\Psi}-\langle \mathcal{O}_i  \rangle_{\lambda_i,\Psi}\langle \mathcal{O}_j \rangle_{\lambda_i,\Psi}=\langle \mathcal{O}_i \mathcal{O}_j \rangle_{\Sigma_{n^*},\Psi}-\langle \mathcal{O}_i  \rangle_{\Sigma_{n^*},\Psi}\langle \mathcal{O}_j \rangle_{\Sigma_{n^*},\Psi}\sim O(N^2).
\eea
Since the state on n-sheeted manifold $\Sigma_n$ is also expected to be dual to some given bulk geometry \cite{Dong:2016fnf}, thus the connected two-point correlator should satisfy  (\ref{correlator_two_point}). The argument can be easily generalized to higher point correlators. Therefore, we conclude that if the state $|\Psi\rangle$ can have a well-defined bulk geometry, the quantum fluctuations in the projection state $\mathbb{P}_\Psi(\lambda_i)|\Psi\rangle$ are suppressed in the semiclassical limit, that is satisfying the  condition (\ref{correlator_two_point}).

\subsection{Superposition of  projection states }\label{section_construction}

In the previous section, we demonstrated that quantum fluctuations are suppressed in the projection state of any given geometric state. This naturally leads to the question of whether projection states themselves can be considered as geometric states. Additionally, new states can be constructed through superposition and mixing of projection states. It is worthwhile to explore the properties of these newly constructed states. 

Actually by defintion the state $|\Psi\rangle$ can be  expressed as superposition state of the projection states.
The state $|\Psi\rangle$ itself can be expressed as
\bea\label{Geometric_decomposition}
|\Psi\rangle =\sum_i \sqrt{\langle \Psi |\mathbb{P}_\Psi(\lambda_i)|\Psi\rangle} |\Phi\rangle_{\lambda_i}=\sum_i \sqrt{\lambda_i \mathcal{P}_\Psi(\lambda_i)}|\Phi\rangle_{\lambda_i}.
\eea 
We could construct the following more general states by superposition with different weight function $g(\lambda_i)$
\bea\label{General_construction_pure}
|\Psi\rangle_g:= \mathcal{N}_g\sum_i g(\lambda_i )\mathbb{P}_\Psi(\lambda_i)|\Psi\rangle, 
\eea
where the normalization constant  
\bea\label{normalization}
\mathcal{N}_g=\left(\sum_i \lambda_i g(\lambda_i)g^*(\lambda_i) \mathcal{P}_\Psi(\lambda_i)\right)^{-1/2}.
\eea
We are interested in the states with holographic dual. If the function $g(\lambda_i)$ is independent with $N$, we can calculate the connected two-point correlation functions of single trace operators $\mathcal{O}_i$ in the new state $|\Psi\rangle_g$. We have
\bea
&&~_g\langle \Psi|\mathcal{O}_i \mathcal{O}_j |\Psi\rangle_g=\mathcal{N}_g^2\sum_i g(\lambda_i) g^*(\lambda_i)\langle \Psi| \mathbb{P}_\Psi(\lambda_i)\mathcal{O}_i \mathcal{O}_j \mathbb{P}_\Psi(\lambda_i)|\Psi\rangle\nn\\
&&\phantom{~_g\langle \Psi|\mathcal{O}_i \mathcal{O}_j |\Psi\rangle_g}= \mathcal{N}_g^2\sum_i \lambda_i g(\lambda_i) g^*(\lambda_i)\mathcal{P}_\Psi(\lambda_i)\langle\mathcal{O}_i \mathcal{O}_j \rangle_{\lambda_i,\Psi}.
\eea
If $|\Psi\rangle_g$ also has a geometric dual, 
we should require the connected two-point correlators scaling as $O(N^2)$. But it is not obvious this condition should be satisfied for general function $g(\lambda_i)$. Let us consider the case $g=1$, that is the state $|\Psi\rangle$. We would have
\bea
&&\langle \Psi|\mathcal{O}_i \mathcal{O}_j |\Psi\rangle-\langle \Psi|\mathcal{O}_i|\Psi\rangle \langle \Psi| \mathcal{O}_j |\Psi\rangle\nn \\
&&=\sum_i \lambda_i \mathcal{P}_\Psi(\lambda_i)\langle\mathcal{O}_i \mathcal{O}_j \rangle_{\lambda_i,\Psi}-\sum_{i'}\lambda_{i'} \mathcal{P}_\Psi(\lambda_{i'})\langle\mathcal{O}_{i}  \rangle_{\lambda_{i'},\Psi}\sum_{i''}\lambda_{i''} \mathcal{P}_\Psi(\lambda_{i''})\langle\mathcal{O}_{i}  \rangle_{\lambda_{i''},\Psi}\nn \\
&&=\sum_i \lambda_i \mathcal{P}_\Psi(\lambda_i)\langle\mathcal{O}_i\rangle_{\lambda_i,\Psi} \langle \mathcal{O}_j \rangle_{\lambda_i,\Psi}-\sum_{i',i''}\lambda_{i'} \lambda_{i''}\mathcal{P}_\Psi(\lambda_{i'})\mathcal{P}_\Psi(\lambda_{i''})\langle\mathcal{O}_{i}  \rangle_{\lambda_{i'},\Psi} \langle\mathcal{O}_{i}  \rangle_{\lambda_{i''},\Psi}+O(N^2),\nn
\eea
where we have used (\ref{Projection_factor}). The condition (\ref{correlator_two_point}) gives the constraint
\bea\label{constraint_twopoint}
\sum_i \lambda_i \mathcal{P}_\Psi(\lambda_i)\langle\mathcal{O}_i\rangle_{\lambda_i,\Psi} \langle \mathcal{O}_j \rangle_{\lambda_i,\Psi}-\sum_{i',i''}\lambda_{i'} \lambda_{i''}\mathcal{P}_\Psi(\lambda_{i'})\mathcal{P}_\Psi(\lambda_{i''})\langle\mathcal{O}_{i}  \rangle_{\lambda_{i'},\Psi} \langle\mathcal{O}_{i}  \rangle_{\lambda_{i''},\Psi}\lesssim O(N^2).
\eea
Since we have $\langle \mathcal{O}_i\rangle_{\lambda_i,\Psi} \sim O(N^2)$, the above constraint is non-trivial. Actually it suggests the density of eigenvalues for the geometric states should satisfy some specific property. In \cite{Guo:2020roc} we have shown there exists an approximated state $|\Phi\rangle_{\lambda_0}$ with $t_0:=-b^\Psi-\log \lambda_0=S(\rho_A^\Psi)-b^\Psi$ for general geometric state $|\Psi\rangle$. This means the expectation value of any local operator $\mathcal{O}$ approaches to $tr(\rho_A^\Psi \mathcal{O})$ in the large $N$ limit. By using (\ref{Geometric_decomposition}) this suggests the function $\lambda_i \mathcal{P}_\Psi(\lambda_i)$ would approach to $\delta(\lambda_i-\lambda_0)$ in the limit $N\to \infty$. Or more precisely in the large $N$ limit we have
\bea\label{spectra_delta}
 \int d\lambda \lambda \mathcal{P}_\Psi(\lambda) f(\lambda) \to f(\lambda_0),
\eea
for any function $f(\lambda)$. We can write (\ref{constraint_twopoint}) by integration. Applying (\ref{spectra_delta}) to the integration, we find the order $O(N^4)$ term of the left hand side of the constraint (\ref{constraint_twopoint}) must be vanishing. Considering that there exists $O(N^{-2})$ correction to $\lambda_i \mathcal{P}(\lambda_i)$, the left hand of the constraint (\ref{constraint}) may be of order $O(N^2)$. In Appendix.\ref{section_fluctuation_vacuumprojection} we use the projection state of vacuum to show the results.


For general state $|\Psi\rangle_g$ we can obtain the constraint
\bea\label{constraint_general}
&&\mathcal{N}_g^2\sum_i \lambda_i |g(\lambda_i)|^2\mathcal{P}_\Psi(\lambda_i)\langle\mathcal{O}_i\rangle_{\lambda_i,\Psi} \langle \mathcal{O}_j \rangle_{\lambda_i,\Psi}\nn\\
&&-\mathcal{N}_g^4\sum_{i',i''}\lambda_{i'} \lambda_{i''}|g(\lambda_{i'})|^2|g(\lambda_{i'})|^2\mathcal{P}_\Psi(\lambda_{i'})\mathcal{P}_\Psi(\lambda_{i''})\langle\mathcal{O}_{i}  \rangle_{\lambda_{i'},\Psi} \langle\mathcal{O}_{i}  \rangle_{\lambda_{i''},\Psi}\lesssim O(N^2).\nn\\
~
\eea
By using (\ref{normalization}) and (\ref{spectra_delta}) we have
\bea
\mathcal{N}_g \to |g(\lambda_0)|^{-1},
\eea
in large $N$ limit. Again by using (\ref{spectra_delta}) it can be shown the left hand side of (\ref{constraint_general}) is vanishing in the limit $N\to \infty$. Therefore, the general states $|\Psi\rangle_g$ do satsify the constraints for the geometric states. The argument can be easily generalized to higher point correlation functions.  

With the pure states one could the mixed states, which can be generally written as
\bea
\rho_{m}= \sum_I P_I |\Psi\rangle_{g_I} ~_{g_I}\langle \Psi|, 
\eea
where $P_I\in (0,1)$ and $\sum_I P_I=1$. 

In Section \ref{section_universal} we study the expectation value of $T(w)$ in the projection states. We find that a universal divergent term exists near the boundary of $A$. For the newly constructed state $|\Psi\rangle_g$, the expectation value of $T(w)$ may also be divergent. However, one may obtain a state for which the expectation value of $T(w)$ is smooth near the boundary $x'$, provided that $g(\lambda_i)$ satisfies the following conditions
\bea
&&\sum_i|g(\lambda_i)|^2\left(\lambda_i \mathcal{P}(\lambda_i)+\int_{\lambda_{i}}^{\lambda_m}d\lambda \log\frac{\lambda_i}{\lambda}\mathcal{P}(\lambda) \right)=0, \label{constraint_general_state}\\
&&\sum_i|g(\lambda_i)|^2\mathcal{P}_{x'}(\lambda_i)=0.
\eea 
It would be interesting to find the solution of the above constraints.

\subsection{Fixed area state}\label{section_fixed}

The projection state (\ref{projection_state_norm}) exhibits interesting properties akin to those of the so-called fixed area state constructed in \cite{Akers:2018fow} and \cite{Dong:2018seb} using the gravitational path integral. In this section, we aim to explore the relationship between these two states. Define the noramlized projection state 
\bea\label{projection_geometry}
|\Phi\rangle_{\lambda_i}:= \frac{\mathbb{P}_\Psi(\lambda_i)|\Psi\rangle}{\sqrt{\langle \Psi |\mathbb{P}_\Psi(\lambda_i)|\Psi\rangle}}.
\eea
The reduced density matrix of the projection state $|\Phi\rangle_{\lambda_i}$ is given by
\bea
\rho_A^{\lambda_i}=\frac{\mathbb{P}_\Psi(\lambda_i)}{\mathcal{P}_{\Psi}(\lambda_i)}.
\eea
$\rho_A^{\lambda_i}$ is propotional to the projector, thus it has flat spectra. One could evaluate the EE, the result is $S(\rho_A^{\lambda_i})=\log \mathcal{P}_\Psi(\lambda_i)$. By definition we also have
\bea
(\rho_A^{\lambda_i})^n=\frac{\mathbb{P}_\Psi(\lambda_i)}{\mathcal{P}_\Psi(\lambda_i)^n},
\eea
thus the R\'enyi entropy is
\bea\label{Renyi_projection}
S_n(\rho_A^{\lambda_i})=S(\rho_A^{\lambda_i})= \log \mathcal{P}_\Psi(\lambda_i),
\eea
which is independent with  the index $n$. It appears that the projection state shares similar properties with the fixed area state. However, as demonstrated below, the projection state $\mathbb{P}|(\lambda_i)|0\rangle$ cannot be considered equivalent to the fixed area state. We will also illustrate how to construct the correct state by combining multiple projection states through superposition.


We would like to show the fixed area state can be constructed by $|\Psi\rangle_g$ (\ref{General_construction_pure}). Let us choose the function $\chi(\lambda_i)$ as
\bea\label{chifunction}
\chi(\lambda)=\begin{cases} 
1 & \text{if } \lambda \in (\lambda_i,\lambda_i+\Delta \lambda_i) , \\
0 & \text{others},
\end{cases}
\eea
with $\Delta \lambda_i/\lambda_i \ll 1$.
Thus, $|\Psi\rangle_\chi$ can be considered the state that belongs to the subspace of eigenstates with eigenvalues in the range $(\lambda_i,\lambda_i+\Delta\lambda_i)$, which is expressed as
\bea\label{chi_geometry}
|\Psi\rangle_\chi=\mathcal{N}_\chi \sum_{\lambda \in (\lambda_i,\lambda_i+\Delta \lambda_i)}\mathbb{P}_\Psi(\lambda)|\Psi\rangle,
\eea 
where
\bea
\mathcal{N}_\chi\simeq\left(\lambda_i \mathcal{P}_\Psi(\lambda_i)\Delta\lambda_i\right)^{-1/2}.
\eea
The reduced density matrix for $|\Psi\rangle_\chi$ is 
\bea\label{reduced_chi}
\rho_A^\chi=\frac{\sum_{\lambda \in (\lambda_i,\lambda_i+\Delta \lambda_i)}\mathbb{P}_\Psi(\lambda)}{\mathcal{P}_\Psi(\lambda_i)\Delta \lambda_i}.
\eea
$\rho_A^\chi$ still has flat spetra, the R\'enyi entropy and EE is given by
\bea
S_n(\rho_A^\chi)=S(\rho_A^\chi)=\log \left(\mathcal{P}_\Psi(\lambda)\Delta \lambda_i \right).
\eea
Now if we require $\Delta\lambda_i =  \lambda_i\Delta_i$, where $\Delta_i\ll 1$ is some constant of order $O(N^0)$. We have
\bea\label{Renyi_chi}
S_n(\rho_A^\chi)=S(\rho_A^\chi)=\log \left(\lambda_i\mathcal{P}_\Psi(\lambda_i) \right)+\log \Delta_i.
\eea
Eq.(\ref{lambdaPlambda_geometry}) shows $\lambda_i \mathcal{P}_\Psi(\lambda_i)\sim O(e^{1/G})\sim O(e^{N^2})$, thus the term $\log \Delta_i$ can be ignored in the large $N$ limit. 

\subsubsection{The holographic dual}\label{section_fixed_correct}
In this section we would like to discuss the bulk metric dual to the projection state $|0\rangle_{\lambda_i}$ defined by (\ref{projection_geometry}) and $|0\rangle_\chi$ defined by (\ref{chi_geometry}) with $|\Psi\rangle=|0\rangle$. In Section \ref{section_stress_vacuum} we calculate the expectation value of $T(w)$ and $\bar T(\bar w)$  in the projection state (\ref{expectation_vacuum_norm}). The result can be easily generalized to the state $|0\rangle_{\chi}$. We have
\bea
&&~_{\chi}\langle 0|T(w)|0\rangle_{\chi}\nn \\
&&=\mathcal{N}_{\chi}^2\sum_{\lambda\in(\lambda_i,\lambda_i+\Delta \lambda_i)} \lambda tr(\mathbb{P}(\lambda)T(w))\nn \\
&&\simeq  \mathcal{N}_{\chi}^2 \lambda_i tr(\mathbb{P}(\lambda_i)T(w))\Delta\lambda_i\nn \\
&&=\frac{tr(\mathbb{P}(\lambda_i)T(w))}{\mathcal{P}(\lambda_i)}=\langle T(w)\rangle_{\lambda_i,0}.
\eea
Thus the expectation value of $T(w)$ is same for the two states at the leading order. Given the expectation value of $T(w)$ and $\bar T(\bar w)$ one could construct the bulk metric in AdS$_3$, as reviewed in Appendix \ref{section_appendix_metric}. However, it appears that the two different states $|0\rangle_{\lambda_i}$ and $|0\rangle_\chi$ are dual to the same bulk metric. This is not too surprising, as metrics and states in CFTs are not in one-to-one correspondence. Metrics can only capture the classical properties of the gravity theory. For example, one could introduce an $O(1/G)$ perturbation to a given bulk geometry, such as a black hole. In principle, the boundary state should also change. However, the backreaction of this perturbation on the metric is negligible in the semiclassical limit. Thus, the bulk metric can effectively describe both states. The differences between $|0\rangle_{\lambda_i}$ and $|0\rangle_\chi$ are significant, and other probes can be used to distinguish them. We will demonstrate that only $|0\rangle_\chi$ can be dual to the bulk metric given by (\ref{geometrydual_projection}), as shown in Appendix \ref{section_appendix_metric}.

Let us now determine which state correctly corresponds to the bulk metric (\ref{geometrydual_projection}). The unnormalized states $$|\tilde{0}\rangle_{\lambda_i}=\mathbb{P}(\lambda_i)|0\rangle$$ and $$|\tilde{0}\rangle_{\chi}=\sum_{\lambda\in (\lambda_i,\lambda_i+\Delta\lambda_i)}\mathbb{P}(\lambda)|0\rangle$$ can be prepared using the Euclidean path integral in the field theory. The normalization of these states is determined by the path integral over the boundary theory. Through the AdS/CFT correspondence, this path integral can be translated into a bulk gravitational path integral. In the field theory, one can evaluate the normalization as follows
\bea
&&~_{\lambda_i}\langle \tilde{0}|\tilde{0}\rangle_{\lambda_i}=\lambda_i \mathcal{P}(\lambda_i)=\frac{\sqrt{b}I_1(2\sqrt{-b(b+\log\lambda_i)})}{\sqrt{-b-\log\lambda_i}}\propto e^{2\sqrt{b t_i}},\nn \\
&&~_{\chi}\langle \tilde{0}|\tilde{0}\rangle_{\chi}\simeq\lambda_i\Delta\lambda_i\mathcal{P}(\lambda_i)=\lambda_i \Delta_i\frac{\sqrt{b}I_1(2\sqrt{-b(b+\log\lambda_i)})}{\sqrt{-b-\log\lambda_i}}\propto e^{-b-t_i+2\sqrt{bt_i}},\nn
~
\eea
where in the last steps we have used the large $c$ limit and the fact that $I_n(x)\propto e^{x}$ in large $x$ limit. Thus, the two states yield different partition functions. According to the AdS/CFT correspondence, the partition function $Z$ on the boundary is related to the on-shell action of the bulk solution in the semiclassical limit  $G\to 0$, that is $Z\simeq e^{-I}$. For the given bulk metric (\ref{geometrydual_projection}), the on-shell action is given by $I(\mathcal{M}_1)$ (\ref{onshell_action}) with $n=1$. It is clear that the state $|\tilde{0}\rangle_\chi$ provides the correct result.

Another check  involves the R\'enyi entropy. Eq.(\ref{Holographic_Renyi_fixed}) gives the holographic R\'enyi entropy. The R\'enyi entropy for the state $|0\rangle_{\lambda_i}$ is given by (\ref{Renyi_projection}), in the large $c$ limit we have
\bea\label{Renyi_projection_lambda}
S_n(\rho_A^{\lambda_i})\simeq (\sqrt{b}+\sqrt{t}_i)^2.
\eea
While the R\'enyi entropy for the state $|0\rangle_{\chi}$ is given by
\bea\label{Renyi_chi}
S_n(\rho_A^{\chi})\simeq 2\sqrt{b t_i},
\eea
which gives the expected result in the holographic side (\ref{Holographic_Renyi_fixed}).

The above results are consistent with those in \cite{Guo:2021tzs}. In that paper, we use the parameterization $\lambda_i=e^{-b-t_i}$, thus $\Delta \lambda_i =\lambda_i \Delta t_i$, which aligns with the above discussions. In this parameterization, the density of eigenvalues is $\mathcal{P}'(t_i)=\lambda_i\mathcal{P}(\lambda_i)$, where $\mathcal{P}'(t_i)$ is the deisnty of eigenstates used in \cite{Guo:2021tzs}.
\subsubsection{Fixed area states in general case}
In the previous section, we analyzed the holographic dual of the state $|0\rangle_{\chi}$. For a general state $|\Psi\rangle$, one can also construct the state $|\Psi\rangle_{\chi}$ (\ref{chi_geometry}), which shares similar properties with the fixed area states.

For the general case, one could obtain the R\'enyi entropy for the state $\Psi\rangle_\chi$ by using (\ref{reduced_chi}). The result is
\bea
S_n(\rho_A^\chi)\simeq \log(\Delta\lambda_i\mathcal{P}(\lambda_i))\simeq \frac{\mathcal{B}_{n^*}}{4G},
\eea
where we have used (\ref{lambdaPlambda_geometry}) and $\Delta\lambda_i=\lambda_i \Delta_i$.
Note that the right hand side is independent of the index $n$, $n^*(\lambda_i)$ only depends on the parameter $\lambda_i$. Moreover, $\mathcal{B}_{n^*}$ is the area of the cosmic brane with the tension $\mu_{n^*}=\frac{n^*-1}{4G n^*}$. The results are consistent with the fixed area states constructed in \cite{Dong:2018seb} via the gravitational path integral, as detailed in \cite{Guo:2021tzs}. We expect the states $|\Psi\rangle_\chi$ can be considered candidates for the fixed area states. Of course, we can also construct unitary equivalent states
\bea
|\Psi'\rangle_\chi:= U_A\otimes U_{\bar A}|\Psi\rangle_\chi,
\eea
where $U_{A(\bar A)}$ is unitary operator located in $A(\bar A)$. It is obvious the R\'enyi entropy of $A$ is same with $|\Psi\rangle_\chi$. One could distinguish these states by evaluating the expectation value of local operators. 

With the construction of $|\Psi\rangle_\chi$ one could decompose the pure state $|\Psi\rangle$ as the superposition of $|\Psi\rangle_{\chi_I}$ for a series of functions $\chi_I$. We can choose the index $I$ to indicate the interval $(\lambda_i,\lambda_i+\Delta \lambda_i)$, $\Delta \lambda_i$ can be chosen small enough to ensure no overlap between different intervals.  $\chi_I$ is defined as (\ref{chifunction}), satisfying $\chi_I \chi_{I'}=\delta_{I,I'}\chi_I$ and $\sum_I \chi_I=1$ over the interval $(0,\lambda_m)$. Thus we have
\bea\label{superposition_fixed}
|\Psi\rangle=\sum_I \mathcal{N}_{\chi_I}|\Psi\rangle_{\chi_I}.
\eea
Based on the fixed area state one could construct the area operator $\hat{A}$, for which $|\Psi\rangle_{\chi_I}$ are the eigenstates. We will provide further comments in the discussion section and conduct a detailed study of the area operator in the near future.

\section{Spectra for the non-Hermtion transition matrix}\label{section_nonhermitian}
In previous sections, we mainly focused on the density matrices, whose spectra lie in the interval $[0,1]$. Recently, there has been significant research on non-Hermitian transition matrices $\mathcal{T}$, which can be seen as a generalization of density matrices \cite{Nakata:2020luh}. Similarly, for a system with a division into $A$ and $\bar A$, one could define the reduced transition matrix by 
\bea
\mathcal{T}_A:=tr_{\bar A}\mathcal{T}.
\eea
In general, the reduced transition matrix $\mathcal{T}_A$ is also non-Hermitian. Thus the eigenvalues of it would also be complex generally. One could define the entanglement measure for $\mathcal{T}_A$, such as the pseudo-R\'enyi entropy
\bea\label{pseudoRenyi}
S_n(\mathcal{T}_A):=\frac{\log tr\mathcal{T}_A^n}{1-n}.
\eea
In the limit $n\to 1$ we obtain the so-called pseudoentropy, $S(\mathcal{T}_A)=-tr \mathcal{T}_A\log \mathcal{T}_A $. For example, given any given pure state $|\psi\rangle$ and $|\phi\rangle$ with $\langle \phi |\psi\rangle\ne 0$, the transition matrix is defined as
\bea\label{transition_matrix_general}
\trans=\frac{|\psi\rangle \langle \phi|}{\langle \phi|\psi\rangle}.
\eea 

For the operator $\mathcal{T}_A$ the spectrum $\sigma(\mathcal{T}_A)$ is a set on the complex plane $\mathbb{C}$. Thus
the density of eigenvalues is defined as
\bea
\rho(x,y):=\sum_i \delta(x- Re \lambda_i)\delta(y-Im \lambda_i),
\eea
which is a function depending on coordinates on the two-dimensional complex plane. Or using the complex coordinate $z=x+i y$ and $\bar z=x-i y$ we have
\bea
\rho(z,\bar z)=\sum_i \delta(z-\lambda_i)\delta(\bar z-\bar \lambda_i).
\eea
In random matrix theory, we are also interested in the eigenvalue density. The usual approach is to consider the Stieltjes transform of the given matrix. Similarly, we can use this method to obtain the density of eigenvalues for the operator $\mathcal{T}_A$. The resolvent of $\mathcal{T}_A$ is $R_{\mathcal{T}_A}:=\frac{1}{z-\mathcal{T}_A}$. To evaluate $\rho(z,\bar z)$ one could define the Green's function of transition matrix $\mathcal{T}_A$\footnote{Given an $N\times N$ random matrix $M$, the Green's function is usually defined as $G(z,\bar z)=\frac{1}{N}tr \frac{1}{z-M}$. Here our definition is slightly different. }
\bea
G(z,\bar z;\mathcal{T}_A):= \sum_i \frac{\lambda_i}{z-\lambda_i}=tr \left(R_{\mathcal{T}_A}\mathcal{T}_A\right).
\eea
By using the following relation
\bea
\partial_{\bar z} \frac{1}{z}=\pi \delta(x)\delta(y)=\pi \delta(z)\delta(\bar z),
\eea
we can obtain the density of eigenvalues 
\bea\label{spetra_derivative}
z\rho(z,\bar z)=\frac{1}{\pi}\partial_{\bar z}G(z,\bar z).
\eea
In Appendix \ref{section_appendix_spectra} we show how to obtain the density of eigenvalues for the density matrix of the  example discussed in Section \ref{section_Density_interval}. 

We can also use the Riesz projection to obtain the projector for an eigenvalue and density of eigenvalue. For an eigenvalue $\lambda_i$ we can define the projection similar as (\ref{Riesz_projection}),
\bea\label{Riesz_projection_TA}
\mathbb{P}_{\mathcal{T}_A}(\lambda_i)=\frac{1}{2\pi i} \int_{\gamma^i}dz R_{\mathcal{T}_A}(z),
\eea 
where $\gamma^i$ is contour that encloses only the point $\lambda_i$.
The density of eigenvalues $\lambda_i$ can be obtained by
\bea\label{density_nonhermitian}
\lambda_i \mathcal{P}(\lambda_i)=\frac{1}{2\pi i} \int_{\gamma^i}dz G(z,\bar z)=\frac{1}{2\pi i} \int_{\gamma^i}dz \sum_{n=1}^\infty \frac{tr\mathcal{T}_A^n}{z^n}.
\eea
Thus, according to previous discussions one could obtain the density of eigenvalues by using $tr\mathcal{T}_A^n$, which is related to the pseudo-R\'enyi entropy by the defintion (\ref{pseudoRenyi}). However, evaluating the pseudo-R'enyi entropy in QFTs is challenging, and there are currently only a few analytical results available. In the following, we will use some examples to demonstrate how to evaluate the density of eigenvalues $\mathcal{P}(\lambda_i)$.

\subsection{Pseudo-Hermitian transition matrix}
For the general transition matrix (\ref{transition_matrix_general}) the eigenvalues are expected to be complex. Thus the  pseudo-R\'enyi entropy would also be complex. In  \cite{Guo:2022jzs} the authors consider a special class of transition matrix, which is $\eta$-pseudo-Hermitian 
\bea
{\trans}^\dagger =\eta \trans \eta^{-1},
\eea
where $\eta$ is an invertible Hermitian operator. Further, it is shown that if the operator $\eta$ can be factorized as $\eta=\eta_A\otimes \eta_{\bar A}$ with both $\eta_A$ and $\eta_{\bar A}$ being invertible and   Hermitian, the eigenvalues of $\mathcal{T}_{A(\bar A)}$ would be real or complex coming with conjugated pairs.  In this case we find $\mathcal{T}_A$ is also pseudo-Hermitian, that is $\mathcal{T}_A^\dagger= \eta_A \mathcal{T}_A \eta_A^{-1}$. 
If $\eta_{A(\bar A)}$ is positive operator, it can be shown that the transition matrix $\mathcal{T}_A$ can be mapped to a density matrix by
\bea
\rho_{A}=\eta_A^{1/2}\mathcal{T}_A\eta_A^{-1/2}. 
\eea
Thus the Green function for $\mathcal{T}_A$ would be same as the Green function for $\rho_A$. By using (\ref{density_nonhermitian}) we can conclude that $\mathcal{T}_A$ has same eigenvalues as $\rho_A$, and the density of eigenvalues are also same. By using (\ref{Riesz_projection_TA}) the projection would also be related by
\bea
\mathbb{P}_{\mathcal{T}_A}=\eta_A^{1/2} \mathbb{P}_{\rho_A}\eta_A^{-1/2}.
\eea
It is obvious that $\mathbb{P}_{\mathcal{T}_A}$ is non-Hermitian. 

\subsection{Local excitation in 2D CFTs}\label{section_local}
Another example is the local excitation state considered in \cite{Nakata:2020luh} \cite{Guo:2022jzs} \cite{Guo:2023aio}. For 2-dimensional CFTs, let us consider the transition matrix
\bea\label{local_transition}
\trans=\frac{\mathcal{O}(w_1,\bar w_1)|0\rangle \langle 0|\tilde{\mathcal{O}}(w_2,\bar w_2)}{\langle \tilde{\mathcal{O}}(w_2,\bar w_2)\mathcal{O}(w_1,\bar w_1) \rangle },
\eea
where $\mathcal{O}$ and $\tilde{\mathcal{O}}$ are two different primary operators. One could evaluate the pseudo-R\'enyi entropy $S_n(\transA)$ by using replica method, which is  related to the $2n$-point correlation function $n$-sheet Riemann surface. But for general theory it is hard to evaluate $S_n(\transA)$. We will consider the operators in 2D massless free boson theory, 
\bea
&&\mathcal{O}(w_1,\bar w_1)=:e^{\frac{i}{2}\phi(0,-a)}:+e^{i\theta}:e^{-\frac{i}{2}\phi(0,-a)}:,\nn \\
&&\tilde{\mathcal{O}}(w_2,\bar w_2)=:e^{\frac{i}{2}\phi(0,-a)}:+:e^{-\frac{i}{2}\phi(0,-a)}:,
\eea
where $w_1=-i a$ and $w_2=i a$, $\theta\in (0,2\pi)$. It is obvious that the transition matrix (\ref{local_transition}) is non-Hermitian if $\theta \ne 0$. We can consider the Lorentzian time evolution by analytical continuation $a\to \delta +i t$ and $a'\to \delta -it$, where $\delta$ is the UV cutoff. Let us choose the subsystem $A=[x_1,x_2]$. In \cite{Guo:2023aio} the authors show the pseudo-R\'enyi entropy can be obtained by using quasiparticle picture and sum rule for pseudo-R\'enyi entropy. The result is
\bea
S_n(\transA)=S_n(\rho_{0,A})+\Delta S_n,
\eea 
where $\rho_{0,A}$ is the reduced density matrix for vacuum state, and
\bea
\Delta S_n=\frac{1}{1-n}\log \frac{1+e^{in\theta}}{(1+e^{i\theta})^n}
\eea
By definition we have
\bea\label{tracelocal}
tr(\transA)^n= e^{-nb +\frac{b}{n}} \frac{1+e^{in\theta}}{(1+e^{i\theta})^n}.
\eea
By using (\ref{density_nonhermitian}) we have
\bea\label{quasiparticle_density}
&&\lambda_i \mathcal{P}(\lambda_i)=\frac{1}{2\pi i} \int_{\gamma^i}dz \sum_{n=1}^\infty \frac{tr(\transA)^n}{z^n}\nn \\
&&\phantom{\lambda_i \mathcal{P}(\lambda_i)}=\frac{1}{2\pi i} \int_{\gamma^i}dz \sum_{k=0}^\infty \frac{b^k}{k!}\left(Li_k(\frac{1}{ze^b(1+e^{i\theta})})+Li_k(\frac{1}{ze^b(1+e^{-i\theta})}) \right)\nn \\
&&\phantom{\lambda_i \mathcal{P}(\lambda_i)}=\frac{1}{2\pi i} \int_{\gamma^i}dz \int_0^\infty dt \frac{\sqrt{b}I_1(2\sqrt{bt})}{\sqrt{t}}\left(\frac{\frac{e^{-b-t}}{1+e^{i\theta}}}{z-\frac{e^{-b-t}}{1+e^{i\theta}}}+\frac{\frac{e^{-b-t}}{1+e^{-i\theta}}}{z-\frac{e^{-b-t}}{1+e^{-i\theta}}}\right).\nn
\eea
The above result suggests the eigenvalues $\lambda_i=\frac{e^{-b-t_i}}{1+e^{i\theta}}$ and $\lambda_i^*=\frac{e^{-b-t_i}}{1+e^{-i\theta}}$, where $t_i\in (0,+\infty)$. Thus, the eigenvalues are complex but come in conjugate pairs. This result is consistent with the expectation that the transition matrix in this example is pseudo-Hermitian \cite{Guo:2022jzs}. By evaluating the integral (\ref{quasiparticle_density}), we find that the density of eigenvalues for $\lambda_i$ and $\lambda_i^*$ are same, which is given by
\bea
\mathcal{P}(\lambda_i)=\mathcal{P}(\lambda_i^*)=\frac{\sqrt{b}I_1(2\sqrt{b t_i})}{e^{-b-t_i}\sqrt{t_i}}.
\eea
Again this is also consistent with the expectation that the transition matrix is pseudo-Hermitian.

As a check of the above result we can evaluate the pseudo-R\'enyi entropy by the definition 
\bea
&&tr(\transA)^n=\sum_i \lambda_i^n +\sum_i (\lambda_i^*)^n\nn \\
&&\phantom{tr(\transA)^n}=\frac{1+e^{in\theta}}{(1+e^{i\theta})^n}\sum_i e^{-n b-n t_i}\nn \\
&&\phantom{tr(\transA)^n}=\frac{1+e^{in\theta}}{(1+e^{i\theta})^n}e^{-nb +\frac{b}{n}},
\eea
which is same with (\ref{tracelocal})\footnote{One could evaluate the sum $\sum_i e^{-nb-nt_i}$ by integration $\int_0^\infty dt e^{-nb-n t}\mathcal{P}(e^{-b-t})$, where $\mathcal{P}(e^{-b-t})$ is the density of eigenvalues for the vacuum. Here we should consider the contribution from the maximal eigenvalue, which is a delta function.}.

From the above results, one can gain more insight into the quasiparticle picture for a locally excited state. The eigenvalues of the transition matrix are given by the product of the eigenvalues $e^{-b-t_i}$ of $\rho_{0,A}$  and $(\frac{1}{1+e^{i\theta}}, \frac{1}{1+e^{-i\theta}})$. This suggests the reduced transition matrix $\mathcal{T}_A$ can be effectively expressed as
\bea
\mathcal{T}_A=\rho_A\otimes \mathcal{R}_A,
\eea 
with $$\mathcal{R}_A:=\left(
\begin{array}{cc}
 \frac{1}{1+e^{i\theta}} & 0 \\
 0 & \frac{1}{1+e^{-i\theta}} \\
\end{array}
\right),$$
which can be seen as the reduced transition matrix for the quasiparticle.
 
\subsection{Sum rule for the Green's function}\label{section_sumrule}
In \cite{Guo:2023aio} and \cite{Guo:2024edr} the authors find there exists sum rules for the pseudo-R\'enyi entropy and R\'enyi entropy of superposition state. Consider the transition matrix (\ref{transition_matrix_general}). To introduce the sum rule we define the superposition state
\bea
|\xi(\theta)\rangle:=\mathcal{N}(\theta)(|\psi\rangle +e^{i\theta}|\phi\rangle,
\eea
where $\theta\in (-\pi,\pi)$ and $\mathcal{N}(\theta)=1/\sqrt{2+e^{i\theta}\langle \phi|\psi\rangle+e^{-i\theta}\langle\psi| \phi\rangle}$ is the normalization constant. Then we can define the reduced density matrix $\rho_A(\theta)=tr_{\bar A} |\xi(\theta)\rangle \langle \xi(\theta)|$. The sum rule is given by
\bea\label{sumrule}
(\transA)^n=\frac{1}{2\pi}\int_{-\pi}^\pi d\theta e^{-i n\theta}\mathcal{N}(\theta)^{-2n}\langle \phi| \psi\rangle^{-n} \rho_A(\theta)^n.
\eea
We can define the Green's function for $\rho_A(\theta)$ as $G(z,\bar z; \rho(\theta)):=tr(\rho(\theta)R_{\rho(\theta)})$
Then the sum rule (\ref{sumrule}) gives the relation between two Green functions
\bea
G(z,\bar z;\mathcal{T}_A)=\frac{1}{2\pi}\int_{-\pi}^\pi d\theta e^{-i n\theta}\mathcal{N}(\theta)^{-2n}\langle \phi| \psi\rangle^{-n} G(z,\bar z; \rho(\theta)).
\eea
For the example in  Section \ref{section_local} one could construct the superposition state $|\xi(\theta)\rangle$ and check the relation for two Green's functions.

\section{Summary and Discussion}\label{section_summary}

In this paper we study the spectral projection of density matrix in QFTs. For any compact operator $M$, we construct the projection operator $\mathbb{P}_M(\lambda_i)$ by using Riesz projection formula (\ref{Riesz_projection}), where $\lambda_i$ is the eigenvalue of $M$. As briefly mentioned in Section \ref{section_density}, the Hilbert space of QFTs is generally infinite, and even the local reduced density matrix may not be well-defined. Recently, it has been shown that this problem can be addressed using the tool of crossed product from operator algebra  \cite{Witten:2021unn},\cite{Chandrasekaran:2022cip}-\cite{AliAhmad:2024eun}. It has also been demonstrated that a covariant regulator for entanglement entropy (EE) can be found based on the crossed product approach \cite{Kudler-Flam:2023hkl}. In our approach in this paper, the projections already incorporate the UV cut-off, as we have used the regularized R\'enyi entropy. Investigating whether projectors are associated with this regularization by crossed product would be an interesting direction for future research.

Our primary focus is on the spectral projection  $\mathbb{P}_\rho(\lambda_i)$ operator for a given density matrix $\rho$ in QFTs. We demonstrate how to evaluate the density of eigenvalues $\mathcal{P}(\lambda_i)$ by using the information from R\'enyi entropy. Furthermore, using $\mathbb{P}_\rho(\lambda_i)$ we construct the projection state $|\phi\rangle_{\lambda_i}$ (\ref{projection_state_norm}), which can be interpreted as a maximally entangled state between $A$ and $\bar A$. The correlation functions of operators $\mathcal{O}_A$ located in $A$ can be computed in the projection state $|\phi\rangle_{\lambda_i}$. We provide an example to show that the density of eigenvalues $\mathcal{P}(\lambda_i)$ and correlator $\langle \mathcal{O}_A\rangle_{\lambda_i,\psi}$ (\ref{expectation_lambda}) are consistent with the definitions (\ref{continuous}) and (\ref{Expectation_alt}) used in \cite{Guo:2020roc}.

We have provided two different methods to evaluate the interesting quantities associated with the eigenstates at eigenvalue $\lambda_i$. From a mathematical perspective, the previous approach is associated with the inverse Laplace transformation, while the spectral projection method in this paper is related to contour integration on the complex plane of eigenvalues $\lambda$. Each method has its advantages and disadvantages. It appears challenging to obtain the density of eigenvalues for the maximal eigenvalue using the spectral projection method, whereas the inverse Laplace transformation method makes this straightforward. The inverse Laplace transformation method is also very useful for analyzing cases associated with holographic states, as it involves the saddle point approximation, an application demonstrated in Section \ref{Fluctuations_projection}. Conversely, the spectral projection method cannot be used in this scenario. However, spectral projections are very useful for expressing projection states and constructing new states through superposition and mixing, as shown in Section \ref{section_construction}. Moreover, the spectral projection method can be easily generalized to non-Hermitian operators, such as transition matrices, as briefly discussed in Section \ref{section_nonhermitian}. In contrast, the inverse Laplace transformation method is not applicable to the non-Hermitian case, where the spectra are generally complex.

In this paper, we identify a universal divergent term for the expectation value of the stress-energy tensor $T(w)$ in the projection state $|\phi\rangle_{\lambda_i,\psi}$. The results are detailed in Section \ref{section_universal}. We find that the expectation value of $T(w)$ is divergent near the boundary  with the coordinate $x'$. The leading divergent term is of order $O((w-x')^{-2})$, with coefficients dependent only on the eigenvalue $\lambda_i$ and density of eigenvalues $\mathcal{P}(\lambda_i)$.  The next divergent term is of order $O((w-x')^{-1})$, with coefficients depending only on the function $\mathcal{P}_{x'}(\lambda_i):=\sum_i \frac{\partial \lambda_i}{\partial x'}\delta(\lambda-\lambda_i)$, which is studied in \cite{Guo:2023tys}. These results can also be derived using the inverse Laplace transformation, as shown in Section \ref{section_universal_ILM}.

Given the projection states, one can construct new states through superposition and mixing. We briefly discuss the properties of these states, focusing on their holographic aspects. It is generally expected that the quantum fluctuations of single-trace operators should be suppressed in the large $N$ limit, satisfying the condition (\ref{correlator_two_point}). Using the inverse Laplace transformation method and saddle point approximation, we show that the correlators in the projection state indeed satisfy the constraint (\ref{Projection_factor}) when the state $|\Psi\rangle$ meets the condition (\ref{correlator_two_point}). By superposition, one can construct the general state (\ref{General_construction_pure}), which also satisfies the condition (\ref{constraint_general}). This implies that quantum fluctuations in these states are suppressed, suggesting that they can also have a well-defined bulk geometry dual. However, as we illustrate in the following example, the projection state $\mathbb{P}(\lambda_i)|0\rangle$ is not expected to correspond to a bulk metric. Nonetheless, by suitably superposition of appromiate amount of projection states, we construct the state $|0\rangle_{\chi}$ (\ref{chi_geometry}), which produces the correct bulk partition function and R\'enyi entropy. We will discuss the implications of this result in a later section. 

Finally, we provide examples demonstrating that the spectral projection method can be applied to non-Hermitian operators, particularly transition matrices. The application of this method to non-Hermitian operators will be explored further in future work.

In this paper, we conduct foundational studies on spectral projection. However, there are many topics that remain unexplored. We will briefly outline potential applications based on the findings of this paper.

\subsection{More properties of the spectral projection}

In Section \ref{section_correlator_local}, we examine the expectation value of local operators $\mathcal{O}_A$ in the projection state $|\phi\rangle_{\lambda_i,\psi}$. This can be evaluated using the expectation value of the local operators on the $n$-sheeted surface $\Sigma_n$. Additionally, one can consider the expectation value of operators that involve both $\mathcal{O}_A$ and $\mathcal{O}_{\bar A}$, where $\mathcal{O}_{A(\bar A)}$ are operators located in region $A(\bar A)$,  respectively. That is
\bea\label{correlator_AbarA}
\langle \psi| \mathbb{P}(\lambda_i) \mathcal{O}_A \mathcal{O}_{\bar A}\mathbb{P}(\lambda_i)|\psi\rangle.
\eea 
These correlators provide insight into the correlation between the regions  $A$ and $\bar A$, thereby offering more information about the wave function of the projection states. However, the correlator (\ref{correlator_AbarA}) is significantly more challenging to compute than the correlators in (\ref{expectation_value}). To evaluate the correlators in (\ref{correlator_AbarA}), it appears necessary to consider the off-diagonal elements of the eigenstates, which may be obtained by employing the method used in \cite{Brehm:2018ipf}.

In this paper, we focus exclusively on the projection states of a given pure state $|\psi\rangle$. One could also explore the overlap between the spectral projections of two different states. 
Given two different pure state $|\psi\rangle$ and $|\phi\rangle$, one could obtain the reduced density matrices $\rho_A^\psi= tr_{\bar A}|\psi\rangle \langle \psi|$ and $\rho_A^\phi= tr_{\bar A}|\phi\rangle \langle \phi|$. By using the Riesz projection (\ref{Riesz_projection}) one can obtain the projectors for eigenvalue $\lambda_i$
\bea
&&\mathbb{P}_\psi(\lambda_i)=\frac{1}{2\pi i} \int_{\gamma^i}dz \frac{1}{z-\rho_A^\psi},\nn \\
&&\mathbb{P}_\phi(\lambda_i)=\frac{1}{2\pi i} \int_{\gamma^i}dz' \frac{1}{z'-\rho_A^\phi}.
\eea
Let us consider the projection $\mathbb{P}_\psi(\lambda_i)$ working on $\rho^\phi_A$, that is
\bea
&&tr (\mathbb{P}_\psi(\lambda_i)\rho_A^\phi)\nn \\
&&=\frac{1}{2\pi i} \int_{\gamma^i}dz tr\left(\frac{1}{z-\rho_A^\psi}\rho_A^\phi\right)\nn \\
&&=\frac{1}{2\pi i} \int_{\gamma^i}dz \sum_{n=1}^\infty \frac{tr[(\rho_A^\psi)^n \rho_A^\phi]}{z^n}.
\eea
To obtain the above result we should evaluate  $tr[(\rho_A^\psi)^n \rho_A^\phi]$, which is discussed in \cite{Lashkari:2015dia} to calculate the relative entropy. In certain cases, analytical results can be obtained, as demonstrated in many papers, for example \cite{Sarosi:2016oks} \cite{He:2017txy}.

One could also consider the overlap between two different projections, which is given by:
\bea
&&tr_A(\mathbb{P}_\psi(\lambda_i)\mathbb{P}_\phi(\lambda_i'))\nn \\
&&=\left(\frac{1}{2\pi i}\right)^2 \int_{\gamma^i}dz \int_{{\gamma^i}'}dz' \sum_{n=1,m=1}^\infty \frac{tr[(\rho_A^\psi)^n (\rho_A^\phi)^m]}{z^n {z'}^m}.
\eea
In QFTs one could evaluate $tr[(\rho_A^\psi)^n (\rho_A^\phi)^m]$ by using replica method.

\subsection{Contruction of non-Hermitian transition matrices}

In Section \ref{section_construction}, we demonstrate how to construct more general states through the superposition of projection states  $\mathbb{P}_\Psi(\lambda_i)|\Psi\rangle$. These projection states form a linear vector space  $\mathcal{H}_P$, within which the state state $|\Psi\rangle_g$ can be considered an element. Additionally, one can introduce an inner product within this space
\bea
~_g\langle\Psi| \Psi\rangle_{g'}= \mathcal{N}_g\mathcal{N}_{g'}\sum_i \lambda_i\mathcal{P}_\Psi(\lambda_i)g(\lambda_i)g'(\lambda_i).
\eea
Thus $\mathcal{H}_P$ is a Hilbert space. 

Within $\mathcal{H}_P$ we can construct the non-Hermitian transition matrices, such as
\bea\label{nonhermitian_transition}
\mathcal{T}^{g|g'}:=\frac{|\Psi\rangle _{g'}~_g\langle\Psi| }{~_g\langle\Psi| \Psi\rangle_{g'}}.
\eea
It is obvious that $\mathcal{T}^{g|g'}$ is non-Hermitian if $g\ne g'$. In this example we could easily compute the pseudo-R\'enyi entropy and pseudoentropy, providing a new framework to explore the properties of non-Hermitian transition matrices. It would be particularly interesting to discuss the holographic dual of non-Hermitian transition matrices and the holographic pseudoentropy.

Similar to density matrices, one can compute the connected two-point correlation functions in transition matrices, as discussed in Section \ref{Fluctuations_projection}. We expect that these transition matrices will also satisfy the constraint (\ref{constraint_general}), indicating that quantum fluctuations are suppressed. This suggests that transition matrices may correspond to a well-defined bulk geometry. It would be intriguing to explore the exact duality between the boundary transition matrix (\ref{nonhermitian_transition}) and the bulk metric.

Moreover, in \cite{Guo:2023aio} and \cite{Guo:2024edr}, the authors identify a sum rule for the reduced transition matrix and density matrix of superposition states, which we briefly discuss in Section \ref{section_sumrule}. Considering the transition matrix mentioned above, it is interesting to note that the superposition state
\bea
|\xi(\theta)\rangle\propto |\Psi\rangle _{g'}+e^{i\theta} |\Psi\rangle _{g},
\eea
may also have a bulk geometric dual. It is worthwhile to investigate the significance of the sum rule in the context of bulk geometry.

\subsection{Superposition and geometry}

In Section \ref{section_fixed} we explore the potential holographic duals of the projection states $|\Phi\rangle_{\lambda_i,\Psi}$ and the superposition states $|\Psi\rangle_\chi$. We conclude that the projection states $|\Phi\rangle_{\lambda_i,\Psi}$ cannot yield a correct bulk geometry dual. To obtain accurate results for the partition function and holographic R\'enyi entropy, it appears necessary to use an appropriate amount of superposition of projection states.

In \cite{Bao:2019fpq}\cite{Bao:2018pvs}, for a given geometric state $|\Psi\rangle$ the authors use the tensor networks to construct an approximate state, which is given by
\bea\label{approximate}
|\Psi'\rangle=\sum_{I=0}^{e^{O(\sqrt{S})}}\sum_i^{e^{S-O(\sqrt{S})}}\sqrt{\lambda_I}|I,i\rangle_A |I,i\rangle_{\bar A}, 
\eea 
where $S$ is the EE of subsystem $A$ in the state $|\Psi\rangle$. For a fixed $I$ we have a  maximal entangled states with the dimension $e^{S-O(\sqrt{S})}$, which actually are related to the projection states $\mathbb{P}_\Psi(\lambda_i)|\Psi\rangle$. 

Consider the vacuum state as an example. 
The EE of projection state $|0\rangle_{\lambda_i}$ is given by $S(\rho_A^{\lambda_i})\simeq (\sqrt{b}+\sqrt{t_i})^2$ with $\lambda_i=e^{-b-t_i}$ (\ref{Renyi_projection_lambda}). While the EE of the state  $|0\rangle_{\chi}$ is $S(\rho_A^\chi)\simeq 2\sqrt{b t_i}$ (\ref{Renyi_chi}). For the vacuum state we have $S(\rho_A)=2b$. If $t_i=b$ we would have $S(\rho_A^\chi)=S(\rho_A)$. The approximate state $|0'\rangle$ for vacuum state $|0\rangle$ can be obtained by superposition of  $|0\rangle_\chi$ with the amount of order $e^{O(\sqrt{S})}$. Eq.(\ref{superposition_fixed})  suggests one could express the state $|0\rangle$ as superposition of a series of state $|0\rangle_{\chi_I}$. Meanwhile, Equation (\ref{approximate}) indicates that it is possible to construct an approximate state by superposing a limited number of $|0\rangle_{\chi_I}$. It would be interesting to investigate how to choose the set $\chi_I$ to achieve this.

A more general question is whether the state $|\Psi\rangle_g$ (\ref{General_construction_pure}), constructed by superposing projection states, can have a well-defined geometric dual. We plan to explore this question further in the near future.

~
\\~
~\\
\textbf{Acknowledgements}\\

I would like to thank Song He, Jie-qiang Wu and Jiaju Zhang for their valuable discussions. I am also grateful to the 2024 Gravitation and Cosmology Workshop held in Datong, where parts of this draft were presented.I am supposed by the National Natural Science Foundation of China under Grant No.12005070 and the Fundamental Research Funds for the Central Universities under Grants NO.2020kfyXJJS041.

\appendix


\section{Density of eigenvalues by using Eq.(\ref{spetra_derivative})}\label{section_appendix_spectra}

For the non-Hermitian matrix the density of eigenvalue is generally more subtle. Formally, one could compute the density of eigenvalue by using Eq.(\ref{spetra_derivative})   once the Green function $G(z,\bar z)$ is known. In this section we would like to show the equation can be applicable to the density matrix. Let us consider one interval $A$ with length $L$ in vacuum state of 2-dimensional CFTs. The R\'enyi entropy is given by (\ref{Renyivacuum}). 
By definition $G(z,\bar z)$ can be expressed as
\bea
G(z,\bar z)=\sum_{k=0}^{\infty}\frac{b^k}{k!} \sum_{n=1}^\infty \frac{(z e^b)^{-n}}{n^k} = \sum_{k=0}^{\infty}\frac{b^k}{k!}Li_k(\frac{1}{ze^b}),
\eea
where $Li_k(x)$ is the polylogarithm function. The  polylogarithm function can be expressed as an integral form
\bea
Li_k(x)=\frac{1}{\Gamma(k)}\int_0^{\infty}\frac{t^{k-1}}{e^t/x-1}dt,
\eea
for $Re(k)>0$ and all $x$ except for $x\ge 1$. With this we have
\bea
G(z,\bar z)=\frac{1}{ze^b-1}+\int_0^{\infty}dt \frac{\sqrt{b} I_1\left(2 \sqrt{b} \sqrt{t}\right)}{\sqrt{t}(z e^{b+t}-1)}.
\eea

The above formula can also be used to derive the density of eigenvalues for reduced density matrix, for which the eigenvalues are in the region $[0,1]$ and $b$ is positive. Thus we have
\bea
&&z\rho(x,y)=\frac{1}{\pi}\partial_{\bar z}G(z,\bar z) \nn \\
&&=e^{-b} \delta(x-e^{-b})\delta(y)+\int_0^\infty dt \frac{\sqrt{b} I_1\left(2 \sqrt{b} \sqrt{t}\right)}{\sqrt{t}e^{b+t}} \delta(x-e^{-b-t})\delta(y)\nn \\
&&=e^{-b} \delta(x-e^{-b})\delta(y)+H(e^{-b}-x)\frac{b I_1\left(2 \sqrt{b} \sqrt{-b-\log (x)}\right)}{\sqrt{-b (b+\log (x))}}\delta(y),\nn \\~
\eea
which is same with the result in Section \ref{section_Density_interval}. The $\delta(y)$ function in the above expression shows the eigenvalues lie on the $x$-axis.
\section{Details of the calculation of higher point correlators}\label{Appendix_higher_correlator}
In Section \ref{section_higher_order} we discuss the two point correlators of stress energy tensor in the projected state. In this section we would like to show the details of the calculations. One could compute the correlator by using (\ref{expectation_twopoint}).According to (\ref{expectation_twopoint}) we need to evaluate the summation
\bea
\sum_{n=1}^{\infty}\frac{tr[(\rho^\psi_A)^n] }{z^n}\langle T(w)T(w')\rangle_{\Sigma_n}.
\eea
Note that both $\xi$ and $\xi'$ depend on the $n$. We cannot find an analytical expression for the summation. Let us define $a:=\frac{(w-R_0)(R_1-w')}{(R_1-w)(w'-R_0)}$. We have
\bea
\sum_{n=1}^{\infty}\frac{tr[(\rho^\psi_A)^n] }{z^n}\frac{(\zeta \zeta')^{2}}{n^4}\frac{c/2}{(\zeta-\zeta')^4}=\sum_{n=1}^{\infty}\frac{tr[(\rho^\psi_A)^n] }{z^n} \frac{1}{n^4}\frac{c/2}{(a^{1/2n}-a^{-1/2n})^4}.
\eea
Assume the distance between $w$ and $w'$ is small, that is $a\ll 1$. We have
\bea
&&\sum_{n=1}^{\infty}\frac{tr[(\rho^\psi_A)^n] }{z^n} \frac{1}{n^4}\frac{c/2}{(a^{1/2n}-a^{-1/2n})^4}\nn \\
&&=\frac{c}{2}\sum_{n=1}^{\infty}\frac{e^{b/n} }{(z e^b)^n}\frac{1}{n^4}\left(\frac{2 n^4}{(a-1)^3}+\frac{n^4}{(a-1)^4}+\frac{\left(n^2-1\right) n^2}{6 (a-1)}+\frac{\left(7 n^2-1\right) n^2}{6 (a-1)^2}+O((a-1)^0) \right)\nn \\
&&=\frac{c}{12}\sum_{k=0}^{\infty}\frac{b^k}{k!}\frac{a}{(a-1)^4}  \left[ (a^2+4a+1) \text{Li}_k\left(\frac{1}{z e^b}\right)-(a-1)^2 \text{Li}_{k+2}\left(\frac{1}{z e^b}\right)\right]+O((a-1)^0)\nn \\
&&=\frac{c}{12} \left[\frac{a(a^2+4a+1)}{(a-1)^4} \int_0^\infty dt \frac{e^{-b-t}}{z-e^{-b-t}}\frac{\sqrt{b} I_1(2\sqrt{bt})}{\sqrt{t}}-\frac{a}{(a-1)^2}\int_0^\infty dt \frac{e^{-b-t}}{z-e^{-b-t}}\frac{\sqrt{t} I_1(2\sqrt{bt})}{\sqrt{b}} \right]\nn \\
&&\ \ \ +O((a-1)^0).
\eea
Thus, we obtain
\bea
&&\frac{1}{2\pi i} \int_{\gamma^i} dz \sum_{n=1}^{\infty}\frac{tr[(\rho^\psi_A)^n] }{z^n} \frac{1}{n^4}\frac{c/2}{(a^{1/2n}-a^{-1/2n})^4}\nn \\
&&=\frac{c}{12} \left[\frac{a(a^2+4a+1)}{(a-1)^4} \frac{\sqrt{b} I_1(2\sqrt{b t_i})}{\sqrt{t_i}}-\frac{a}{(a-1)^2}\frac{\sqrt{t_i} I_1(2\sqrt{bt_i})}{\sqrt{b}} \right]+O((a-1)^0),\nn
\eea
where $t_i=-b-\log \lambda_i$. We will also need to deal with the following summation
\bea
&&\sum_{n=1}^{\infty}\frac{tr[(\rho^\psi_A)^n]}{z^n}\left(\frac{c}{24}\frac{n^2-1}{n^2}\right)^2\nn \\
&&=\frac{c^2}{576}\sum_{n=1}^{\infty}\frac{e^{b/n}}{(ze^b)^n}\left(\frac{n^2-1}{n^2}\right)^2\nn \\
&&=\frac{c^2}{576} \sum_{k=0}^\infty \frac{b^k}{k!}\sum_{n=1}^\infty \frac{1}{(ze^b)^n n^k}\left(\frac{n^2-1}{n^2}\right)^2\nn \\
&&=\frac{c^2}{576} \sum_{k=0}^\infty \frac{b^k}{k!}\left[\text{Li}_k\left(\frac{1}{z e^b}\right)-4 \text{Li}_{k+2}\left(\frac{1}{z e^b}\right)+6 \text{Li}_{k+4}\left(\frac{1}{z e^b}\right)-4 \text{Li}_{k+6}\left(\frac{1}{z e^b}\right)+\text{Li}_{k+8}\left(\frac{1}{z e^b}\right)\right]\nn \\
&&=\frac{c^2}{576}\int_0^\infty dt \frac{e^{-b-t}}{z-e^{-b-t}} \Big[ \frac{\sqrt{b} I_1\left(2 \sqrt{b} \sqrt{t}\right)}{\sqrt{t}}-\frac{4 \sqrt{t} I_1\left(2 \sqrt{b} \sqrt{t}\right)}{\sqrt{b}}+\frac{6 t^{3/2} I_3\left(2 \sqrt{b} \sqrt{t}\right)}{b^{3/2}}\nn \\
&&\phantom{\frac{c^2}{576}\int_0^\infty dt \frac{e^{-b-t}}{z-e^{-b-t}} \Big[}-\frac{4 t^{5/2} I_5\left(2 \sqrt{b} \sqrt{t}\right)}{b^{5/2}}+\frac{t^{7/2} I_7\left(2 \sqrt{b} \sqrt{t}\right)}{b^{7/2}}\Big].
\eea
Thus we obtain
\bea
&&\frac{1}{2\pi i} \int_{\gamma^i} dz\sum_{n=1}^{\infty}\frac{tr[(\rho^\psi_A)^n]}{z^n}\left(\frac{c}{24}\frac{n^2-1}{n^2}\right)^2\nn \\
&&=\frac{c^2}{576} \Big[ \frac{\sqrt{b} I_1\left(2 \sqrt{bt_i}\right)}{\sqrt{t_i}}-\frac{4 \sqrt{t_i} I_1\left(2 \sqrt{bt_i}\right)}{\sqrt{b}}+\frac{6 t_i^{3/2} I_3\left(2 \sqrt{bt_i}\right)}{b^{3/2}}\nn \\
&&\phantom{=\frac{c^2}{576} \Big[}-\frac{4 t_i^{5/2} I_5\left(2 \sqrt{bt_i}\right)}{b^{5/2}}+\frac{t_i^{7/2} I_7\left(2 \sqrt{bt_i}\right)}{b^{7/2}}\Big],
\eea
where $t_i=-b-\log \lambda_i$.

\section{Fluctuations in the projection state of vacuum state}\label{section_fluctuation_vacuumprojection}

In Section \ref{section_construction} we construct the new states $|\Psi\rangle_g$ by projection states and smearing functions $g(\lambda_i)$. To make the new states have geometric dual, we would expect the quantum fluctuation would be suppressed in the semiclassical limit $G\to$0. Thus some constraints are necessary, for example Eq.(\ref{constraint}). In this section we would like to use the projection states for the vacuum state to show the constraints are satisfied. In 2D CFTs we have $c\sim N^2$. We would like to choose the single trace operator to be $T(w)$. It is obvious that $\langle T(w)T(w')\rangle-\langle T(w)\rangle \langle T(w')\rangle\sim O(c)\sim O(N^2)$. Eq.(\ref{twopointcorrelator_T}) shows that (\ref{Projection_factor}) is satisfied for $T(w)$.

Now we can check the expected constraints (\ref{constraint}). We can express the summation  by integration
\bea\label{int1}
&&\sum_i \lambda_i \mathcal{P}(\lambda_i)\langle T(w)\rangle_{\lambda_i,0} \langle \mathcal T(w') \rangle_{\lambda_i,0}-\sum_{i',i''}\lambda_{i'} \lambda_{i''}\mathcal{P}(\lambda_{i'})\mathcal{P}(\lambda_{i''})\langle T(w)  \rangle_{\lambda_{i'},0} \langle T(w')  \rangle_{\lambda_{i''},0}\nn \\
&&=\int_0^{\lambda_m} d\lambda \lambda \mathcal{P}(\lambda)\langle T(w)\rangle_{\lambda,0} \langle \mathcal T(w') \rangle_{\lambda,0}-\int_0^{\lambda_m} d\lambda \lambda \mathcal{P}(\lambda)\langle T(w)\rangle_{\lambda,0} \int_0^{\lambda_m} d\lambda \lambda \mathcal{P}(\lambda)\langle T(w')\rangle_{\lambda,0}.\nn\\
~
\eea
Using the expression (\ref{expectation_vacuum_norm}) for $\langle T(w)\rangle_{\lambda_i,0}$. We just need to compute 
\bea
&&\int_0^{\lambda_m} d\lambda \lambda \mathcal{P}(\lambda)\left(1-\frac{-b-\log\lambda}{b} \right)^2-\left(\int_0^{\lambda_m} d\lambda \lambda \mathcal{P}(\lambda)\left(1-\frac{-b-\log\lambda}{b} \right) \right)^2\nn \\
&&=\frac{2}{b}+O(e^{-c}).
\eea
Thus the integration (\ref{int1}) is given by
\bea
\frac{2}{b}\left(\frac{c}{24}\right)^2\frac{(R_1-R_0)^2}{(w-R_0)^2(R_1-w)^2}\frac{(R_1-R_0)^2}{(w'-R_0)^2(R_1-w')^2}\sim O(c)\sim O(N^2),
\eea
which is consistent with Eq.(\ref{constraint}).

Let us also check the expected relation (\ref{spectra_delta}). Changing the variable $\lambda=e^{-b-t}$, we have
\bea
\int_{0}^{\lambda_m} d\lambda \lambda \mathcal{P}(\lambda)f(\lambda)=\int_0^\infty dt e^{-2b-2t}\mathcal{P}(e^{-b-t})f(e^{-b-t}).
\eea
Using (\ref{density}) we have
\bea
e^{-2b-2t}\mathcal{P}(e^{-b-t})=e^{-b-t}\frac{\sqrt{b}I_1(2\sqrt{bt})}{\sqrt{t}}\to \delta(t-b)(1-\frac{3}{8}\frac{1}{2b}+...),
\eea
where we use the fact $I_1(z)=\frac{e^{z}}{\sqrt{2\pi  z}}(1-\frac{3}{8}\frac{1}{z}+...)$ in the large $z$ limit, in the last step we take the large $c$ limit. Using the above formula the relation (\ref{spectra_delta}) can be shown. We further see the $O(c^{-1})$ or $O(N^{-2})$ correction to the delta function. One could also check the constraint  (\ref{constraint_general}) for the general states $|\Psi\rangle_g$ as above .

\section{Metric dual to the fixed area state}\label{section_appendix_metric}

Given the expectation value of $T(w)$ and $\bar T(\bar w)$ we can write down the metric in the bulk 
\bea\label{geometrydual_projection}
ds^2=\frac{dy^2}{y^2}+\frac{L_{\lambda_i}}{2}dw^2+\frac{\bar L_{\lambda_i}}{2}d\bar w^2+\left(\frac{1}{y^2}+\frac{y^2}{4}L_{\lambda_i} \bar L_{\lambda_i}\right)dwd\bar w,\nonumber\\
~
\eea
where $L_{\lambda_i}:=-\frac{12}{c}\langle T(w)\rangle_{\lambda_i,0}$, $\bar L_{\lambda_i}:=-\frac{12}{c}\langle \bar T(w)\rangle_{\lambda_i,0}$. Taking (\ref{expectation_vacuum_norm}) into (\ref{geometrydual_projection}), one could obtain the bulk metric. Now we have to meet with a problem that the two states $|0\rangle_{\lambda_i}$ and $|0\rangle_{\chi}$ would have same bulk geometry dual. We will show in Section \ref{section_fixed_correct} the correct dual state of the bulk metric should be $|0\rangle_\chi$.

Let us firstly consider the physics quantities computed in the bulk. It is also noted in \cite{Guo:2021tzs} that the above metric can be mapped to the Poincar\'e coordinate 
\bea
ds^2=\frac{d\eta^2+ d\xi d\bar \xi}{\eta^2},
\eea
by coordinate transformation. The corresponding boundary conformal transformation is given by (\ref{conformal}). One could use this to evaluate the holographic EE for the subsystem $A=[R_0,R_1]$. The result is
\bea
S_A=\frac{\alpha c}{3}\log \frac{R_1-R_0}{\epsilon}=2\sqrt{b t_i},
\eea
where $\alpha=\sqrt{\frac{t_i}{b}}$. Given the bulk geometry, one could use gravitational path integral  and replica method to evaluate the R\'enyi entropy.  We have
\bea
tr\rho_A=\frac{Z_n}{Z_1^n},
\eea  
where $Z_n$ denotes the $n$-replica bulk path integral obtained by replica method. In semi-classical limit the path integral is given by
\bea
Z_n\simeq e^{-I(\mathcal{M}_n)},
\eea
where $\mathcal{M}_n$ is the dominant saddle-point bulk solution. For $n=1$, $\mathcal{M}_1$ is given by the metric (\ref{geometrydual_projection}).  The metric for $\mathcal{M}_n$ can be constructed similarly as $\mathcal{M}_1$. In \cite{Guo:2021tzs} we also evaluate the  on-shell action of $\mathcal{M}_n$. We just list the result as below, 
\bea\label{onshell_action}
I(\mathcal{M}_n)=n (1-\alpha^2)b+2\sqrt{b t_{i}}(n\alpha-1)=n(b+t_i)-2\sqrt{b t_i}.
\eea
With this we can evaluate the holographic R\'enyi entropy by
\bea\label{Holographic_Renyi_fixed}
S_n=\frac{I(\mathcal{M}_n)-nI(\mathcal{M}_1)}{n-1}=2\sqrt{b t_i},
\eea
which is independent with index $n$.

\end{document}